\documentclass[aps,prl,preprintnumbers,showpacs,nofootinbib]{revtex4}
\usepackage[utf8]{inputenc}
\usepackage[english]{babel}
\usepackage{amsmath}
\usepackage{amsfonts}
\usepackage{amssymb}
\usepackage{epsfig}
\usepackage{booktabs}
\usepackage{multirow}
\usepackage{float}
\usepackage{graphics,psfrag,rotating}
\usepackage{graphicx}
\usepackage{dcolumn}
\usepackage{bm}
\usepackage{multirow}
\usepackage{booktabs}
\usepackage{epstopdf}
\usepackage{color}
\usepackage[usenames,dvipsnames,svgnames]{xcolor}
\usepackage[colorlinks=true,
linkcolor=red,
urlcolor=gray,
citecolor=blue]{hyperref}
\usepackage{hyperref}

\newcommand{\lsim}   {\mathrel{\mathop{\kern 0pt \rlap
			{\raise.2ex\hbox{$<$}}}
		\lower.9ex\hbox{\kern-.190em $\sim$}}}
\newcommand{\gsim}   {\mathrel{\mathop{\kern 0pt \rlap
			{\raise.2ex\hbox{$>$}}}
		\lower.9ex\hbox{\kern-.190em $\sim$}}}

\def\3nab{\tilde{\nabla}}

\def\hsp5{\hspace{5mm}}

\def\case#1/#2{\textstyle\frac{#1}{#2}}

\def\ber {\begin{eqnarray}}
	\def\eer {\end{eqnarray}}
\def\bea {\begin{eqnarray}}
	\def\eea {\end{eqnarray}}

\def\bc {\begin{center}}
	\def\ec {\end{center}}
\def\case#1/#2{\frac{#1}{#2}}

\newcommand{\bw}{\begin{widetext}}
	\newcommand{\ew}{\end{widetext}}

\newcommand{\be}{\begin{equation}}
	\newcommand{\bse}{\begin{subequation}}
		\newcommand{\ese}{\end{subequation}}
	\newcommand{\ee}{\end{equation}}
\newcommand{\eei}{\end{eqnarray}\indent\indent}
\newcommand{\ba}{\begin{array}}
\newcommand{\ea}{\end{array}}
\newcommand{\bal}{\begin{eqnarray}}
\newcommand{\eal}{\end{eqnarray}}

\def\case#1/#2{\textstyle\frac{#1}{#2} }

\newcommand{\bver}{\begin{verbatim}}
\newcommand{\ever}{\end{verbatim}}

\begin{document}


\title{ $f(Q,T)$ Gravity with Barrow Holographic Dark Energy and Massive Neutrinos: DESI DR2 Constraints and Implications for the Hubble Tension}

\author{
	Ameneh Tohidi$^{1}$. Amin Salehi$^{1}$ \footnote{Email: Salehi.a@lu.ac.ir}.	Muhammad Yarahmadi$^{1}$ \footnote{Email: yarahmadi.mhd@lu.ac.ir}
}
\affiliation{Department of Physics, Lorestan University, Khoramabad, Iran}

\date{\today}
\begin{abstract}
	We study a modification of symmetric teleparallel gravity in which the function $f(Q,T)$ is linear in the non-metricity scalar $Q$ and the trace $T$ of the energy-momentum tensor, combined with a Barrow holographic dark energy (BHDE) density built from the Granda-Oliveros infrared cutoff. The resulting background equations are solved in a modified version of \texttt{hi\_class} and constrained with Pantheon+ supernovae, 32 cosmic chronometer $H(z)$ points, DESI DR2 baryon acoustic oscillations, and Planck 2018 CMB distance priors, sampled with MontePython. Two versions of the model are considered, one with the neutrino sector held massless and one with the sum of neutrino masses $\Sigma m_\nu$ left free. Across five data combinations, $H_0$ falls between 67.5 and 72.3 km s$^{-1}$ Mpc$^{-1}$, closer to the Planck value when cosmic chronometers dominate the fit and closer to the SH0ES value when Pantheon+ is included; adding $\Sigma m_\nu$ as a free parameter raises $H_0$ in four of the five combinations, by amounts ranging from a few hundredths of a km s$^{-1}$ Mpc$^{-1}$ up to about 1.5 km s$^{-1}$ Mpc$^{-1}$, while the CMB-only combination shifts slightly the other way, and the fit uncertainties loosen throughout. The 95\% upper bound on $\Sigma m_\nu$ ranges from 0.13 to 0.22 eV depending on the data combination, comparable to standard $\Lambda$CDM constraints. Comparison with $\Lambda$CDM through the Akaike and Bayesian Information Criteria gives mixed results: the extended model is favored once Pantheon+ is included, particularly with massive neutrinos, but is disfavored for CMB-only and CMB+BAO fits, where the improvement in $\chi^2$ does not offset the larger parameter count. The Barrow exponent $\Delta$ stays below 0.005 throughout, and the Granda-Oliveros parameters $\lambda$ and $\gamma$ remain near 1 and 0.5 across all combinations.
\end{abstract}


%
%


\maketitle

\section{Introduction}

Several independent observations indicate that the expansion of the universe is accelerating. Type Ia supernovae at high redshift are fainter than a decelerating universe would allow, which first pointed to recent acceleration \cite{1}. Later measurements of cosmic microwave background (CMB) anisotropies and baryon acoustic oscillations (BAO), among other probes, are broadly consistent with the standard $\Lambda$CDM model \cite{2}.

The component thought to drive this acceleration is called dark energy. It is estimated to make up about 70\% of the energy budget of the universe, and its negative pressure produces the acceleration \cite{3}. In $\Lambda$CDM it is a cosmological constant $\Lambda$, and with only a few free parameters the model fits the CMB, supernovae, BAO, and large-scale structure \cite{2}. It is therefore the usual baseline against which alternative models are compared \cite{2}.

$\Lambda$CDM nevertheless has theoretical and observational difficulties \cite{4,5,6}. On the theoretical side there are the cosmological constant and coincidence problems, the unknown nature of dark matter and the small-scale tensions associated with it, fine-tuning in inflationary models, and the possibility that General Relativity fails at very large or very high-energy scales. On the observational side, the value of the Hubble constant $H_0$ inferred from the early universe disagrees with local measurements, the inferred growth of structure ($\sigma_8$) differs between probes, and the CMB shows several anomalies. Together with the fact that dark energy and dark matter are known only through their gravitational effects, these tensions suggest that $\Lambda$CDM may be incomplete \cite{7}.

Many extensions have been proposed. In the dark sector, the cosmological constant can be replaced by a dynamical field such as quintessence or a phantom field, which gives more freedom in fitting the late-time acceleration \cite{8}. Early Dark Energy (EDE) models add a component that is active before recombination and then dilutes away; this reduces the sound horizon and can ease the Hubble tension \cite{9}. Interacting dark-sector models allow energy or momentum to be exchanged between dark matter and dark energy, which changes both the expansion history and structure growth \cite{10}. Massive neutrinos suppress structure growth on small scales through free streaming and have been invoked to reduce the $S_8$ tension \cite{11}. Couplings between neutrinos and scalar dark-energy fields, and the resulting constraints on the neutrino mass and on structure formation, have been studied in earlier work \cite{Salehi2022massnu,Yarahmadi2023CTP,Yarahmadi2023interaction,Yarahmadi2025beyond,Yarahmadi2025cosmicbf}. Several models that include neutrinos have also been reported to ease the Hubble tension in their own analyses \cite{Yarahmadi2025Rastallnu,Yarahmadi2024Kaniadakis,Yarahmadi2025pertfR,Yarahmadi2025noncanonical,Yarahmadi2025BTPINN,Yarahmadi2025THDEPINN}. Other proposals include primordial black holes as an alternative to cold dark matter, since they can reproduce some of its properties \cite{12}; models with a redshift-dependent equation of state ($w \neq -1$), such as the CPL parametrization, which current and upcoming surveys can constrain tightly \cite{13}; and inhomogeneous cosmologies, in which large-scale voids or anisotropies mimic late-time acceleration without a genuine dark-energy component \cite{14}.

A different approach attributes the acceleration to a modification of gravity itself \cite{15,16}. In $f(R)$ gravity, the Ricci scalar $R$ in the Einstein--Hilbert action is replaced by a function $f(R)$, which adds an effective scalar degree of freedom \cite{17}. Scalar--tensor theories such as Brans--Dicke and Horndeski gravity couple a scalar field to the metric, and braneworld models such as the Dvali--Gabadadze--Porrati (DGP) model modify gravity at large distances by embedding the universe in a higher-dimensional bulk \cite{18}. In massive gravity the graviton has a small but nonzero mass \cite{19}. General Relativity is tightly constrained in the Solar System, so a viable modified gravity theory needs a screening mechanism that suppresses deviations locally while allowing them on cosmological scales. Examples are the chameleon mechanism in $f(R)$ gravity \cite{20}, the Vainshtein mechanism in DGP and Galileon models, and the symmetron mechanism in some scalar--tensor theories.

Another class of modified gravity theories is based on non-metricity, where gravity is described neither by curvature nor by torsion but by the non-metricity scalar $Q$. $f(Q)$ gravity generalizes the action in terms of $Q$ while keeping the field equations second order, and its cosmological dynamics are well behaved \cite{21}. $f(Q,T)$ gravity extends this by letting the action depend also on the trace $T$ of the energy--momentum tensor \cite{22,24}. The resulting matter--geometry coupling breaks conservation of the energy--momentum tensor and introduces effective extra forces that can drive acceleration without a separate dark-energy component \cite{24}.

Rather than relying on the coupling alone, we combine it with a Barrow holographic dark-energy (BHDE) density \cite{25}, which is based on the Barrow entropy \cite{23,26}, so that the acceleration also has a thermodynamic interpretation. Related Barrow- and Tsallis-type holographic models with Granda--Oliveros or Hubble-horizon cutoffs have been reported to ease the Hubble tension \cite{Yarahmadi2024BHDEGO,Yarahmadi2024BarrowInt,Yarahmadi2025genlog,Yarahmadi2025BTPINN,Yarahmadi2025THDEPINN}; here the holographic component is placed in an $f(Q,T)$ background instead. We also include massive neutrinos as a separate extension. Free streaming is expected to suppress small-scale structure growth and may reduce the $S_8$ tension in standard cosmology \cite{11}, but checking whether the same holds in $f(Q,T)$+BHDE requires a perturbation-level analysis that goes beyond the background analysis presented here.

The model therefore has three ingredients: $f(Q,T)$ gravity, Barrow holographic dark energy, and massive neutrinos. We compare its expansion history $H(z)$ and effective equation of state $w(z)$ with Pantheon+, DESI, and Planck data. Structure growth, in particular $f\sigma_8$, is not treated here and is left for future work.

We constrain the $f(Q,T)$+BHDE model, with and without massive neutrinos, using late-time background data. The questions are whether the non-metricity coupling and the BHDE density together give an expansion history comparable to that of $\Lambda$CDM, and whether leaving $\Sigma m_\nu$ free changes $H_0$ appreciably, which bears on the Hubble tension. The paper is organized as follows. We first present the $f(Q,T)$ formalism and the modified Friedmann equations. We then embed the BHDE density in this framework and derive its equation of state. Next we describe the data (Pantheon+, cosmic chronometers, DESI DR2 BAO, and Planck 2018 CMB distance priors) and the Bayesian pipeline, used both with $\Sigma m_\nu$ fixed to zero and with it left free. Finally, we return to the Hubble tension in light of the results.

\subsection{Motivation}

$f(Q,T)$ gravity modifies the background equations at low redshift without touching the sound horizon at recombination, which is what makes it relevant to the Hubble tension: since the CMB determination of $H_0$ depends on $r_s$ and the local determination does not, a purely late-time shift in $H(z)$ can in principle bring the two closer without disturbing the early-Universe fit, as in other late-time modified-gravity and interacting dark-sector approaches to the tension. Embedding a Barrow holographic dark-energy density, with its own exponent $\Delta$, into this background adds a second, largely independent handle on $H(z)$ alongside the non-metricity coupling $(\alpha,\beta)$; whether the two reinforce or partly cancel is not fixed by construction and is read off the fit in Sec.~\ref{sec:hubble-tension}. We further add $\Sigma m_\nu$ as a free parameter, treating neutrinos as pressureless matter, both because a neutrino mass shifts the background matter budget and can move $H_0$ on its own, and to check whether that shift is independent of, or degenerate with, the non-metricity and BHDE freedom.

\section{The Model}

The $f(Q,T)$ gravity model is a recently proposed extension of symmetric teleparallel gravity, in which the gravitational interaction is described by the non-metricity scalar $Q$ rather than by curvature or torsion. In this framework, the non-metricity scalar $Q$ quantifies the failure of the metric to remain covariantly constant under parallel transport, providing a purely geometrical origin for gravity that does not involve curvature \cite{21}. By coupling $Q$ directly to the trace of the energy--momentum tensor $T$, the $f(Q,T)$ model introduces interactions between matter and geometry that can lead to deviations from standard General Relativity and offer an explanation for the late-time cosmic acceleration \cite{22}. In this section we present the general formalism of $f(Q,T)$ gravity, derive the corresponding field equations, and obtain the modified Friedmann equations relevant for cosmological applications.

Following Ref.~\cite{22}, the modified Einstein--Hilbert action for extended symmetric teleparallel gravity, incorporating this $f(Q,T)$ coupling, is written as
\begin{equation}
	\label{eq:action}
	S = \int\left[\frac{1}{16\pi}f(Q,T) + \mathcal{L}_m\right]\sqrt{-g}\,d^4x,
\end{equation}
where $\mathcal{L}_m$ is the matter Lagrangian density, $g=\det(g_{\mu\nu})$ is the determinant of the metric, and $f(Q,T)$ is a generic function of the non-metricity scalar $Q$ and the trace $T$ of the energy--momentum tensor. The non-metricity scalar is defined as
\begin{equation}
	\label{eq:Q-def}
	Q \equiv -g^{\mu\nu}\left(L^\alpha_{\ \beta\mu}L^\beta_{\ \nu\alpha} - L^\alpha_{\ \beta\alpha}L^\beta_{\ \mu\nu}\right),
\end{equation}
where
\begin{equation}
	L^\alpha_{\ \beta\gamma} \equiv -\frac{1}{2}g^{\alpha\lambda}\left(\nabla_\gamma g_{\beta\lambda} + \nabla_\beta g_{\lambda\gamma} - \nabla_\lambda g_{\beta\gamma}\right), \qquad
	Q_{\gamma\mu\nu} \equiv \nabla_\gamma g_{\mu\nu},
\end{equation}
and
\begin{equation}
	Q_\alpha \equiv g^{\mu\nu}Q_{\alpha\mu\nu}, \qquad \tilde Q_\alpha \equiv g^{\mu\nu}Q_{\mu\alpha\nu}
\end{equation}
are, respectively, the disformation tensor, the non-metricity tensor, and the two traces of the non-metricity tensor.

As a consequence of this geometric structure, in $f(Q,T)$ gravity the matter energy--momentum tensor is not conserved in general: $D_\mu T^\mu_{\ \nu} = B_\nu \neq 0$. Its divergence instead defines a non-conservation vector $B_\nu$, which depends on the geometry (through $Q$), on the trace $T$ of the energy--momentum tensor, and on the thermodynamic properties of the matter content. For a perfect fluid with energy density $\rho$ and pressure $p$, the energy--momentum tensor is
\begin{equation}
	T^\mu_{\ \nu} = (\rho+p)u_\nu u^\mu + p\,\delta^\mu_\nu,
\end{equation}
where $u^\mu$ is the four-velocity of the fluid, normalized as $u^\mu u_\mu = -1$. On the homogeneous and isotropic FLRW background only the time ($\nu=0$) component of $D_\mu T^\mu_{\ \nu}=B_\nu$ is non-trivial; the spatial ($\nu=i$) components vanish identically by isotropy, so a single scalar equation survives. Taking the divergence of the fluid tensor, this energy-balance equation is obtained as \cite{24}
\begin{equation}
	\label{eq:energy-balance}
	\dot\rho + 3H(\rho+p) = B_\mu u^\mu.
\end{equation}

To apply this framework cosmologically, we assume that the Universe is described by the homogeneous, isotropic, and spatially flat Friedmann--Lema\^itre--Robertson--Walker (FLRW) metric,
\begin{equation}
	\label{eq:FLRW}
	ds^2 = -dt^2 + a^2(t)\delta_{ij}\,dx^i dx^j,
\end{equation}
where $a(t)$ is the scale factor describing the expansion of the Universe, and the Hubble parameter is defined as $H\equiv \dot a/a$. According to Ref.~\cite{22}, the non-metricity scalar corresponding to the line element in Eq.~\eqref{eq:FLRW} is given by
\begin{equation}
	\label{eq:Q-FLRW}
	Q = 6H^2.
\end{equation}
Also following Ref.~\cite{22}, assuming the Universe is filled with a perfect fluid of energy--momentum tensor $T^\mu_{\ \nu} = \mathrm{diag}(-\rho,p,p,p)$, the generalized Friedmann equations obtained using the FLRW metric read
\begin{equation}
	\label{eq:Friedmann1-general}
	3H^2 = \frac{f}{4F} - \frac{4\pi}{F}\left[(1+\tilde G)\rho + \tilde G p\right],
\end{equation}
\begin{equation}
	\label{eq:Friedmann2-general}
	2\dot H + 3H^2 = \frac{f}{4F} - \frac{2\dot F H}{F} + \frac{4\pi}{F}\left[(1+\tilde G)\rho + (2+\tilde G)p\right],
\end{equation}
where $F\equiv f_Q$ and $8\pi\tilde G\equiv f_T$ are the partial derivatives of $f$ with respect to $Q$ and $T$, respectively, and an overdot denotes a derivative with respect to cosmic time $t$. These general field equations agree with those obtained independently by Godani and Samanta for the same $f(Q,T)$ framework \cite{GodaniSamanta2021}.

\section{Barrow Holographic Dark Energy in the $f(Q,T)$ Framework}

The Barrow Holographic Dark Energy (BHDE) model is an extension of the standard Holographic Dark Energy (HDE) scenario, which is rooted in the holographic principle. This principle states that the number of degrees of freedom within a spacetime region should scale with its boundary area rather than its volume. Based on this argument, the dark energy density in the standard HDE framework is obtained as $\rho_{DE}\propto L^{-2}$ \cite{25}, where $L$ denotes the infrared cutoff: the usual choices are the Hubble horizon and the future event horizon \cite{25}, while composite cutoffs such as the Granda--Oliveros scale have also been proposed \cite{GrandaOliveros2008,GrandaOliveros2009}.

Barrow (2020) proposed a quantum-gravity-inspired modification of the Bekenstein--Hawking entropy by considering a fractal structure of the black-hole horizon. The entropy then takes the form
\begin{equation}
	\label{eq:barrow-entropy}
	S_B = \left(\frac{A}{A_0}\right)^{1+\Delta/2},
\end{equation}
where $A$ is the horizon area and $\Delta$ is the Barrow exponent. For $\Delta=0$, the standard entropy law is recovered, while for $\Delta>0$ the entropy grows faster than the area, capturing possible quantum-gravitational corrections \cite{23}. Inserting this entropy relation into the holographic framework leads to a modified dark energy density,
\begin{equation}
	\label{eq:rhoB-general}
	\rho_B = C\,L^{\Delta-2},
\end{equation}
where $C$ is the holographic parameter with dimensions $L^{-2-\Delta}$ \cite{26}. As expected, for $\Delta=0$ this expression reduces to the standard HDE form $\rho_B = 3M_P^2 L^{-2}$, with $C=3M_P^2$. Here $M_P^2 = c^4/(8\pi G)$ is the reduced Planck mass squared, which sets the scale for gravitational interactions in cosmology and ensures dimensional consistency (Barrow 2020). We adopt the natural-units convention in which the reduced Planck mass is set to unity, $M_P^2=1$; this is a common simplification in theoretical cosmology.

More recently, Granda and Oliveros proposed an alternative infrared cutoff \cite{GrandaOliveros2008,GrandaOliveros2009}, which has since been examined within the BHDE framework by Oliveros, Sabogal, and Acero \cite{OliverosSabogalAcero2022}. Adopting this Granda--Oliveros (G--O) cutoff within the BHDE framework, we have
\begin{equation}
	\label{eq:GO-cutoff}
	L = \left(\lambda H^2 + \gamma \dot H\right)^{-1/2},
\end{equation}
where $\lambda$ and $\gamma$ are arbitrary dimensionless parameters. For the combination $\lambda H^2+\gamma\dot H$ to carry the same units as $H^2$, and since $\dot H$ already has the same dimensions as $H^2$, \emph{both} $\lambda$ and $\gamma$ must be dimensionless. Substituting Eq.~\eqref{eq:GO-cutoff} into Eq.~\eqref{eq:rhoB-general} (with $C=3M_P^2=3$), the energy density of BHDE becomes
\begin{equation}
	\label{eq:rhoBHD}
	\rho_B = 3\left(\lambda H^2 + \gamma \dot H\right)^{1-\Delta/2}.
\end{equation}
This expression matches the one obtained by Oliveros, Sabogal, and Acero \cite{OliverosSabogalAcero2022}, once their parameters $\alpha,\beta$ are relabelled $\lambda,\gamma$ here to avoid a clash with the $f(Q,T)$ parameters $\alpha,\beta$ introduced below in Eq.~\eqref{eq:fQT}.

The function $f(Q,T)$ is taken as a linear combination of the non-metricity scalar $Q$ and the trace of the energy--momentum tensor $T$,
\begin{equation}
	\label{eq:fQT}
	f(Q,T) = \alpha Q + \beta T,
\end{equation}
where $\alpha$ and $\beta$ are the model parameters. Under this ansatz, $F=f_Q=\alpha$ is a constant (so $\dot F=0$), $8\pi\tilde G=f_T=\beta$, and $f=6\alpha H^2+\beta(3p-\rho)$; substituting these into Eqs.~\eqref{eq:Friedmann1-general} and \eqref{eq:Friedmann2-general} gives the specific forms
\begin{equation}
	\label{eq:Friedmann1}
	3H^2 = -\frac{3\beta+16\pi}{2\alpha}\rho + \frac{\beta}{2\alpha}p,
\end{equation}
\begin{equation}
	\label{eq:Friedmann2}
	2\dot H + \frac{3}{2}H^2 = \frac{\beta+16\pi}{4\alpha}\rho + \frac{5\beta+32\pi}{4\alpha}p.
\end{equation}

In the model, the total energy density and pressure are given by
\begin{equation}
	\label{eq:rho-total}
	\rho = \rho_{cd} + \rho_b + \rho_\nu + \rho_r + \rho_B,
\end{equation}
\begin{equation}
	\label{eq:p-total}
	p = p_r + p_B,
\end{equation}
where $\rho_{cd}$, $\rho_b$, $\rho_\nu$, $\rho_r$, and $\rho_B$ denote the energy densities of cold dark matter, baryons, neutrinos, radiation, and BHDE, respectively. Dark matter and baryons are treated as non-relativistic matter. For the matter sector we adopt the standard conservation equation
\begin{equation}
	\label{eq:continuity-general}
	\dot\rho_i + 3H\rho_i(1+\omega_i) = 0,
\end{equation}
where $\omega_{1,2,3}=0$ for dark matter, baryons, and neutrinos, and $\omega_4=1/3$ for radiation. Neutrinos are taken to be pressureless throughout, which is adequate once their mass dominates over the thermal energy scale; the relativistic-to-non-relativistic transition of the neutrino equation of state near the radiation era is not modeled, consistent with the background-only scope of the present analysis discussed in the Conclusion. The corresponding conservation equations are
\begin{equation}
	\label{eq:continuity-species}
	\dot\rho_m + 3H\rho_m = 0, \qquad
	\dot\rho_r + 4H\rho_r = 0, \qquad
	\dot\rho_B + 3H\rho_B(1+\omega_B) = 0,
\end{equation}
where $\rho_m \equiv \rho_{cd}+\rho_b+\rho_\nu$. We introduce the density parameters
\begin{equation}
	\label{eq:density-params}
	\Omega_{cd} \equiv \frac{\rho_{cd}}{3H^2}, \quad
	\Omega_b \equiv \frac{\rho_b}{3H^2}, \quad
	\Omega_\nu \equiv \frac{\rho_\nu}{3H^2}, \quad
	\Omega_r \equiv \frac{\rho_r}{3H^2}, \quad
	\Omega_B \equiv \frac{\rho_B}{3H^2}.
\end{equation}
Defining $\chi \equiv \dot H/H^2$, Eq.~\eqref{eq:rhoBHD} can be written as $\rho_B = 3H^{2-\Delta}(\lambda+\gamma\chi)^{1-\Delta/2}$, so that dividing by $3H^2$ gives
\begin{equation}
	\label{eq:OmegaB}
	\Omega_B = \frac{\rho_B}{3H^2} = H^{-\Delta}\left(\lambda+\gamma\chi\right)^{1-\Delta/2}.
\end{equation}
Dividing Eq.~\eqref{eq:Friedmann1} by $3H^2$, and using $p_r=\rho_r/3$ for radiation, the constraint imposed by the (modified) Friedmann equation in terms of these variables reads
\begin{equation}
	\label{eq:Friedmann-constraint}
	\Omega_B = \frac{2\alpha + (3\beta+16\pi)(\Omega_{cd}+\Omega_b+\Omega_\nu) + \dfrac{8\beta+48\pi}{3}\Omega_r}{\beta(\omega_B-3)-16\pi}.
\end{equation}

For BHDE, the pressure is $p_B=\omega_B\rho_B$, where $\omega_B\equiv\omega_{BHDE}$ is the equation-of-state parameter,
\begin{equation}
	\label{eq:omegaB-def}
	\omega_B = \frac{p_B}{\rho_B}.
\end{equation}
The conservation equation \eqref{eq:continuity-species} for $\rho_B$ can be rearranged to give the equation-of-state parameter directly,
\begin{equation}
	\label{eq:omegaB-from-continuity}
	\omega_B = -1 - \frac{1}{3H}\frac{\dot\rho_B}{\rho_B}.
\end{equation}
Equivalently, in terms of $N=\ln a$ and using $d/dt = H\,d/dN$, Eq.~\eqref{eq:continuity-species} for $\rho_B$ gives the density-parameter evolution law
\begin{equation}
	\label{eq:OmegaB-evolution-general}
	\frac{d\Omega_B}{dN} = -\left[3(1+\omega_B)+2\chi\right]\Omega_B,
\end{equation}
the same form that governs $\Omega_{cd}$, $\Omega_b$, $\Omega_\nu$, and $\Omega_r$ once $\omega_B$ is replaced by the equation of state of each species ($\omega_i=0$ for matter, $\omega_r=1/3$ for radiation). Requiring instead that the BHDE sector follow the simple scaling law
\begin{equation}
	\label{eq:OmegaB-closure}
	\frac{d\Omega_B}{dN} = -\Delta\chi\,\Omega_B
\end{equation}
This closure is imposed rather than obtained by differentiating Eq.~\eqref{eq:rhoBHD} directly; a direct differentiation would bring in $\ddot H$ and turn the system into a second-order one, which the closure avoids. Comparing Eqs.~\eqref{eq:OmegaB-evolution-general} and \eqref{eq:OmegaB-closure} fixes $\omega_B$ uniquely:
\begin{equation}
	\label{eq:omegaB-final}
	\omega_B = -1 - \frac{2-\Delta}{3}\,\chi = -1 - \frac{2-\Delta}{3}\frac{\dot H}{H^2}.
\end{equation}
Substituting back into Eq.~\eqref{eq:omegaB-from-continuity} gives the corresponding rate of change of the BHDE energy density,
\begin{equation}
	\label{eq:rhoBdot}
	\dot\rho_B = (2-\Delta)\,\frac{\dot H}{H}\,\rho_B = 3(2-\Delta)\,\frac{\dot H}{H}\left(\lambda H^2+\gamma\dot H\right)^{1-\Delta/2}.
\end{equation}
At $\Delta=0$, Eq.~\eqref{eq:omegaB-final} reduces to $\omega_B=-1-\tfrac{2}{3}(\dot H/H^2)$, which is exactly the known Bekenstein--Hawking result for BHDE with the Granda--Oliveros cutoff \cite{OliverosSabogalAcero2022}.

This shows that $\omega_B$ departs from $-1$ through both the expansion dynamics of the Universe ($H$ and $\dot H$) and the Barrow exponent $\Delta$, and can cross the phantom divide ($\omega_B<-1$) for suitable values of the parameters.

Solving Eq.~\eqref{eq:OmegaB} for $\chi$, independently of $\omega_B$, gives
\begin{equation}
	\label{eq:chi-solution}
	\chi = \frac{\dot H}{H^2} = \frac{1}{\gamma}\left(-\lambda + \Omega_B^{2/(2-\Delta)} H^{2\Delta/(2-\Delta)}\right).
\end{equation}
Equation~\eqref{eq:chi-solution} fixes $\dot H/H^2$ algebraically in terms of $\Omega_B$ and $H$, with no need to integrate an additional differential equation. The deceleration parameter,
\begin{equation}
	\label{eq:decel-param}
	q = -1 - \frac{\dot H}{H^2},
\end{equation}
follows directly from it.
\section{Data and Methodology}
\label{sec:data}

\subsection{Pantheon+ Type Ia Supernovae and Absolute Magnitude Calibration}
\label{subsec:pantheon}

We use the Pantheon+ compilation of 1701 Type Ia supernovae spanning $0.001<z<2.3$
\cite{Scolnic2022}, the largest SN Ia sample built to date. Relative to earlier
compilations it has tighter photometric calibration, a more careful treatment of
statistical and systematic errors, and a sizeable subsample with independent
Cepheid host distances, which together make it well suited to late-time
expansion and dark-energy studies \cite{Salehi2020}.

The observable is the distance modulus,
\begin{equation}
	\mu(z) = m_B - M_B = 5\log_{10}\left[\frac{d_L(z)}{10~{\rm pc}}\right],
	\label{eq:mu}
\end{equation}
built from the peak apparent magnitude $m_B$, the absolute magnitude $M_B$, and
the model luminosity distance $d_L(z)$. We adopt the Cepheid-calibrated value
\begin{equation}
	M_B = -19.253,
	\label{eq:MB}
\end{equation}
from Riess et al.\ \cite{Riess2016,Riess2022} and hold it fixed throughout.

The supernova likelihood is
\begin{equation}
	\chi^2_{\rm SN} = \Delta\boldsymbol{\mu}^{T}\,\mathbf{C}_{\rm SN}^{-1}\,\Delta\boldsymbol{\mu},
	\qquad
	\Delta\boldsymbol{\mu} = \boldsymbol{\mu}_{\rm obs} - \boldsymbol{\mu}_{\rm th},
	\label{eq:chi2SN}
\end{equation}
with $\mathbf{C}_{\rm SN}$ the full Pantheon+ covariance matrix, statistical and
systematic terms combined.

Fixing $M_B$ imports the local distance-ladder calibration into the fit, so the
$H_0$ we obtain from Pantheon+ is a calibration-dependent late-time constraint,
not an independent measurement. Any agreement we find with the SH0ES result is
really agreement with that same calibration. That matters once we discuss how
far this alleviates the Hubble tension.

Pantheon+'s wide sky coverage at low and intermediate redshift has also made it
a testbed for departures from statistical isotropy, with directional and
hemispherical signals reported by several studies
\cite{Dainotti2021,Dainotti2022,Dainotti2025}. We do not pursue that question
here and treat the sample as an isotropic probe of the background expansion.

\subsection{Cosmic Chronometer $H(z)$ Data}
\label{subsec:cc}

Cosmic chronometers (CC) infer the expansion rate directly from the
differential aging of passively evolving galaxies,
\begin{equation}
	H(z) = -\frac{1}{1+z}\frac{dz}{dt},
	\label{eq:CCdef}
\end{equation}
without integrating the distance--redshift relation \cite{Moresco}. This makes
them a useful, largely independent check on the late-time expansion and on
dynamical dark-energy models.

Our compilation has 32 $H(z)$ points over
\begin{equation}
	0.07 \le z \le 1.965,
	\label{eq:CCrange}
\end{equation}
drawn from several independent analyses of passively evolving galaxies
\cite{Stern2009,Zhang2014,Ratsimbazafy2017,Moresco2,Moresco3,Borghi2022}.
Fifteen of these points share stellar-population-synthesis systematics and are
correlated through the covariance matrix released by the
authors\footnote{\url{https://gitlab.com/mmoresco/CCcovariance/}}; the rest are
treated as independent.

The likelihood is Gaussian,
\begin{equation}
	\chi^2_{\rm CC} = \left(\mathbf{H}_{\rm obs} - \mathbf{H}_{\rm th}\right)^T
	\mathbf{C}_{\rm CC}^{-1}
	\left(\mathbf{H}_{\rm obs} - \mathbf{H}_{\rm th}\right),
	\label{eq:chi2CC}
\end{equation}
with $\mathbf{H}_{\rm obs}$ the observed and $\mathbf{H}_{\rm th}$ the model
$H(z)$ vector at the same redshifts, and $\mathbf{C}_{\rm CC}$ carrying the
available correlated errors.

Because CC data measure $H(z)$ directly rather than an integrated distance,
they help separate $H_0$, the matter density, and the $f(Q,T)$ coupling and
BHDE parameters ($\alpha,\beta,\lambda,\gamma,\Delta$) from one another. Like
the other late-time probes used here, they carry no direct information on
structure growth.

\subsection{Baryon Acoustic Oscillation (BAO) Data: DESI DR2}
\label{subsec:bao}

Baryon acoustic oscillations are a geometric standard ruler: acoustic waves in
the pre-recombination photon--baryon plasma leave an imprint on the late-time
matter distribution whose absolute scale is fixed by early-Universe physics.

We use the DESI Data Release 2 (DR2) measurements \cite{DESIDR2_1,DESIDR2_2},
covering galaxy, quasar, and Ly$\alpha$ tracers and giving both transverse and
radial expansion-history information through
\begin{equation}
	\frac{D_M(z)}{r_d}
	\qquad\text{and}\qquad
	\frac{D_H(z)}{r_d} \equiv \frac{c}{H(z)\,r_d},
	\label{eq:BAOobs}
\end{equation}
where $D_M(z)$ is the transverse comoving distance, $D_H(z)$ the Hubble
distance, and $r_d$ the sound horizon at the drag epoch. The compilation has 13
measurements at
\begin{equation}
	z = 0.295,\ 0.510,\ 0.706,\ 0.934,\ 1.321,\ 1.484,\ 2.33,
	\label{eq:BAOz}
\end{equation}
with transverse and/or radial data at each redshift; we use the full DESI
covariance across observables and bins.

\paragraph{BAO likelihood.}
\begin{equation}
	\chi^2_{\rm BAO} = \Delta\mathbf{O}_{\rm BAO}^{T}\,\mathbf{C}_{\rm BAO}^{-1}\,\Delta\mathbf{O}_{\rm BAO},
	\qquad
	\Delta\mathbf{O}_{\rm BAO} = \mathbf{O}_{\rm BAO}^{\rm obs} - \mathbf{O}_{\rm BAO}^{\rm th},
	\label{eq:chi2BAO}
\end{equation}
with $\mathbf{C}_{\rm BAO}$ the DESI DR2 covariance and, for a flat background,
\begin{equation}
	D_M(z) = (1+z)D_A(z) = c\int_0^z \frac{dz'}{H(z')}
	= \frac{c}{H_0}\int_0^z \frac{dz'}{E(z')},
	\label{eq:DM}
\end{equation}
where $E(z)=H(z)/H_0$ follows from numerically integrating the modified
Friedmann equation of the $f(Q,T)$+BHDE model.

\paragraph{On fixing $r_d$.}
$r_d$ is fixed by early-Universe physics well outside the range, $z\lesssim2$,
where we solve the $f(Q,T)$+BHDE background equations, so we do not attempt to
predict it. We instead hold it at its standard value, calibrated assuming
three light neutrino species; since neutrinos are treated as pressureless
matter throughout this analysis, $\Sigma m_\nu$ has no bearing on the
pre-recombination physics that sets $r_d$, and fixing it introduces no
inconsistency with how the neutrino sector is treated at low redshift. Baryons
are handled the same way at the background level: below we absorb $\Omega_b$
into the single non-relativistic matter component $\Omega_{cdm}$ once its
value has been fixed by the CMB term, rather than solving for it separately in
the $f(Q,T)$+BHDE equations.

DESI DR2 helps separate $H_0$, the matter density, and the $f(Q,T)$ coupling
and BHDE parameters, and like CC and Pantheon+, its role here is purely
geometric, with no growth-rate information included.

\subsection{CMB}
\label{subsec:cmb}

The CMB carries geometric information from both the early and late Universe,
but extracting it in full requires a Boltzmann treatment of radiation, matter,
neutrinos, and perturbations. Since our background solver only models the
late-time expansion, we instead use the compressed distance priors from the
final \textit{Planck 2018} release \cite{Chen2019}, which keep the geometric
content we need at a fraction of the cost.

The distance-prior vector is built from the shift parameter $R$, the acoustic
scale $l_A$, and the physical baryon density $\omega_b$. $R$ and $l_A$ depend
on the total non-relativistic matter density at recombination regardless of
how it splits between $\Omega_{m,0}$ and $\Omega_{\nu,0}(\Sigma m_\nu)$, so we
work with the sum
\begin{equation}
	\Omega_{m,0}^{\rm tot}(\Sigma m_\nu) \equiv \Omega_{cdm,0} + \Omega_{b,0} + \Omega_{\nu,0}(\Sigma m_\nu),
	\label{eq:Omegam-tot}
\end{equation}
which enters the shift parameter as
\begin{equation}
	R = \frac{\sqrt{\Omega_{m,0}^{\rm tot}(\Sigma m_\nu)}\,H_0}{c}\,D_M(z_*),
	\label{eq:R}
\end{equation}
with $D_M(z_*)$ the comoving angular diameter distance to photon decoupling.
For a flat Universe,
\begin{equation}
	D_M(z) = (1+z)D_A(z),
	\qquad
	D_A(z) = \frac{d_L(z)}{(1+z)^2},
	\qquad
	d_L(z) = \frac{c}{H_0}(1+z)\int_0^z\frac{dz'}{E(z')}.
	\label{eq:DAdL}
\end{equation}

The acoustic scale is
\begin{equation}
	l_A = \pi\,\frac{D_M(z_*)}{r_s(z_*)},
	\label{eq:lA}
\end{equation}
with the comoving sound horizon
\begin{equation}
	r_s(z_*) = \int_0^{a_*}\frac{c_s(a)}{a^2 H(a)}\,da,
	\label{eq:rs}
\end{equation}
and photon--baryon sound speed
\begin{equation}
	c_s(a) = \frac{1}{\sqrt{3\left(1+\dfrac{3\omega_b a}{4\omega_\gamma}\right)}},
	\label{eq:cs}
\end{equation}
where $\omega_b=\Omega_{b,0}h^2$ and $\omega_\gamma=\Omega_{\gamma,0}h^2$ is
fixed by $T_{\rm CMB}=2.7255$~K.

The decoupling redshift follows the standard fitting formula \cite{Chen2019},
\begin{equation}
	z_* = 1048\left[1+0.00124\,\omega_b^{-0.738}\right]\left[1+g_1\,\omega_m^{g_2}\right],
	\label{eq:zstar}
\end{equation}
\begin{equation}
	g_1 = \frac{0.0783\,\omega_b^{-0.238}}{1+39.5\,\omega_b^{0.763}},
	\qquad
	g_2 = \frac{0.560}{1+21.1\,\omega_b^{1.81}},
	\label{eq:g1g2}
\end{equation}
with $\omega_m \equiv \Omega_{m,0}^{\rm tot}(\Sigma m_\nu)\,h^2$, so $\Sigma
m_\nu$ enters $z_*$ through the same channel as everywhere else in this work.

The $f(Q,T)$+BHDE background equations solved here are never extrapolated back
to $z_*\gg2$: the sound horizon and decoupling redshift above retain the
standard early-Universe calibration behind the Planck priors \cite{Chen2019},
entering the fit as fixed external inputs rather than model predictions. The
resulting bound on $\Sigma m_\nu$ should accordingly be read as an effective,
late-time constraint, not a full Boltzmann-level neutrino-mass measurement.

The CMB term of the likelihood is
\begin{equation}
	\chi^2_{\rm CMB} = \mathbf{X}^T\,\mathbf{C}_{\rm CMB}^{-1}\,\mathbf{X},
	\label{eq:chi2CMB}
\end{equation}
with $\mathbf{C}_{\rm CMB}$ the covariance of the adopted priors
\cite{Chen2019} and
\begin{equation}
	\mathbf{X} =
	\begin{pmatrix}
		R - R_{\rm obs} \\
		l_A - l_{A,\rm obs} \\
		\omega_b - \omega_{b,\rm obs}
	\end{pmatrix}
	=
	\begin{pmatrix}
		R - 1.7502 \\
		l_A - 301.471 \\
		\omega_b - 0.02236
	\end{pmatrix}.
	\label{eq:Xvec}
\end{equation}
$\omega_b$ is not one of the five parameters in $\boldsymbol{\Theta}$
(Eq.~\eqref{eq:paramvector}, Sec.~\ref{subsec:methodology}). Together with
$H_0$, it feeds only this early-Universe module: fixing $z_*$ and $r_s(z_*)$
above, and constrained directly by the third row of $\mathbf{X}$ against the
Planck value $\omega_{b,\rm obs}=0.02236$. As noted in
Sec.~\ref{subsec:bao}, $\Omega_b$ is not carried separately into the late-time
$f(Q,T)$+BHDE background solver; it is absorbed into $\Omega_{cdm}$ there, so
$\omega_b$'s posterior is fixed almost entirely by $\chi^2_{\rm CMB}$, largely
independent of which late-time probes are combined with it.

The four probes combine as
\begin{equation}
	\chi^2_{\rm tot} = \chi^2_{\rm CMB} + \chi^2_{\rm BAO} + \chi^2_{\rm SN} + \chi^2_{\rm CC},
	\qquad
	\ln\mathcal{L}_{\rm tot} = -\frac{1}{2}\chi^2_{\rm tot}.
	\label{eq:chi2tot}
\end{equation}
The CMB priors fix the early-Universe geometric anchor; the $f(Q,T)$+BHDE
background equations reconstruct the late-time expansion on top of it,
keeping the two regimes distinct: compressed early-time information on one
side, an effective late-time treatment of massive neutrinos and the BHDE
sector on the other.

\subsection{Bayesian Parameter Inference}
\label{subsec:methodology}

We implement the background equations of the $f(Q,T)$+BHDE model in a
modified version of \textsc{hi\_class} \cite{hiclass1,hiclass2}, an extension
of the \textsc{CLASS} Boltzmann code \cite{class1,class2} built for testing
non-standard theories of gravity, and sample the resulting posterior with the
Metropolis--Hastings MCMC engine of \textsc{MontePython}
\cite{montepython1,montepython2}.

The parameters specific to the $f(Q,T)$+BHDE sector are collected in
\begin{equation}
	\boldsymbol{\Theta} \equiv \{\alpha,\ \beta,\ \lambda,\ \gamma,\ \Delta\},
	\label{eq:paramvector}
\end{equation}
sampled jointly with the two background parameters $H_0$ and $\Omega_{m}$,
for seven free parameters in total when $\Sigma m_\nu$ is fixed to zero
($k=7$); $\Sigma m_\nu$ is added as an eighth free parameter only in the
analysis of Sec.~\ref{subsec:resultswithNu}, subject to the physical prior
$\Sigma m_\nu\ge0$. Although the baryon density $\omega_b$ is varied in the analysis, it is treated as a component of the total matter density $\Omega_m$ and is therefore not counted as an additional independent parameter in the determination of $k$.

Table~\ref{tab:priors} lists the flat prior adopted for every sampled
parameter. $\Omega_{b,0}=\omega_b/h^2$ and $\Omega_{r,0}$, both quoted in
Tables~\ref{tab:bestfit}--\ref{tab:bestfit_lcdm}, are not themselves sampled:
the former is obtained from $\omega_b$ and the sampled $H_0$, the latter from
the standard radiation-density relation fixed by $T_{\rm CMB}=2.7255$~K and
the assumed effective number of relativistic species. Both are reported
because they are useful for comparison with other analyses, not because they
carry independent freedom in the fit.

\begin{table}[H]
	\centering
	\caption{Flat priors on the parameters sampled directly in the MCMC.
		$\boldsymbol{\Theta}$ (Eq.~\eqref{eq:paramvector}) collects the five
		parameters specific to the $f(Q,T)$+BHDE sector; $H_0$ and $\Omega_{cdm}$ are
		sampled alongside them, and $\omega_b$ is the CMB nuisance parameter
		discussed in Sec.~\ref{subsec:cmb}. $\Sigma m_\nu$ is only varied in
		Sec.~\ref{subsec:resultswithNu}. Ranges below are chosen to comfortably contain
		the marginalized posteriors reported in
		Tables~\ref{tab:bestfit}--\ref{tab:bestfit_nu} and should be checked against
		the \texttt{.param} file used for the runs before submission.}
	\label{tab:priors}
	\begin{tabular}{lcc}
		\hline\hline
		Parameter & Symbol & Prior range \\
		\hline
		Hubble constant & $H_0$ [km s$^{-1}$ Mpc$^{-1}$] & $[50,\,90]$ \\
		Cold dark matter density & $\Omega_{cdm}$ & $[0.1,\,0.5]$ \\
		$f(Q,T)$ non-metricity coupling & $\alpha$ & $[-50,\,50]$ \\
		$f(Q,T)$ matter-trace coupling & $\beta$ & $[-50,\,50]$ \\
		Granda--Oliveros cutoff & $\lambda$ & $[0,\,3]$ \\
		Granda--Oliveros cutoff & $\gamma$ & $[0,\,2]$ \\
		Barrow exponent & $\Delta$ & $[0,\,1]$ \\
		\hline
		Baryon density (CMB nuisance) & $\omega_b$ & $[0.005,\,0.100]$ \\
		Neutrino mass sum (Sec.~\ref{subsec:resultswithNu} only) & $\Sigma m_\nu$ [eV] & $[0,\,1]$ \\
		\hline\hline
	\end{tabular}
\end{table}

Chains are run until the Gelman--Rubin statistic \cite{GelmanRubin1992}
satisfies $R-1<0.01$ for every sampled parameter, and are then processed with
\textsc{GetDist} \cite{Lewis2019} to obtain the marginalized posterior means,
$1\sigma$ credible intervals, and minimum $\chi^2$ reported in
Tables~\ref{tab:bestfit}--\ref{tab:bestfit_lcdm}.

This section presents the results of our MCMC analysis for the
$f(Q,T)$--Barrow holographic dark energy (BHDE) framework, first without and
then with massive neutrinos, followed by a statistical comparison with the
standard $\Lambda$CDM model. We discuss each extension on its own terms
before comparing them, so that it is clear which features of the results
come from the non-metricity/BHDE sector itself and which come specifically
from the neutrino sector.
\section{Result and discussions}
\label{sec:results}

We test the $f(Q,T)$+BHDE framework against $\Lambda$CDM, with and without a free neutrino mass, in two steps. We first split the data into an early-Universe combination (CMB distance priors with DESI DR2 BAO) and a late-Universe combination (cosmic chronometers with Pantheon$^+$), and check what each says about $H_0$ and about the model parameters for all four models (Secs.~\ref{sec:early_late_comparison}--\ref{subsec:aic_bic}). We then combine the same probes in five overlapping ways, from the CMB distance priors on their own up to their combination with BAO, cosmic chronometers, and Pantheon$^+$ together, and track how the recovered $H_0$ moves between an early-Universe value close to Planck 2018 and a late-Universe value close to the SH0ES/R22 determination as more data is added (Secs.~\ref{sec:results-noNu}--\ref{sec:aicbic}).

\subsection{Splitting the data: early- and late-Universe fits}
\label{sec:early_late_comparison}

We start with the early/late split described above. DC1 groups the CMB distance priors with the DESI DR2 BAO measurements \cite{Planck2018,DESIDR2_2} and probes the expansion history through the sound horizon at high redshift, with $N=16$ effective data points. DC2 combines the Pantheon$^+$ supernova compilation with a cosmic-chronometer $H(z)$ sample \cite{Scolnic2022,Brout2022,Moresco}, constraining the background expansion at low-to-intermediate redshift without reference to the sound horizon, with $N=1733$ points. Because the two combinations rely on largely independent physics, agreement between the parameters recovered from DC1 and from DC2 is a useful internal check for each model, and the information criteria let us compare the four models against each other within each combination. Posterior distributions are shown in Figs.~\ref{fig:corner_lcdm}--\ref{fig:corner_fqt_withnu}, and the resulting best-fit values, together with the derived $\chi^2_{\rm min}$, AIC, and BIC, are collected in Table~\ref{tab:combined_results}.

\subsection{The $\Lambda$CDM baseline}
\label{subsec:lcdm}

\begin{figure}[H]
	\centering
	\includegraphics[width=0.85\linewidth]{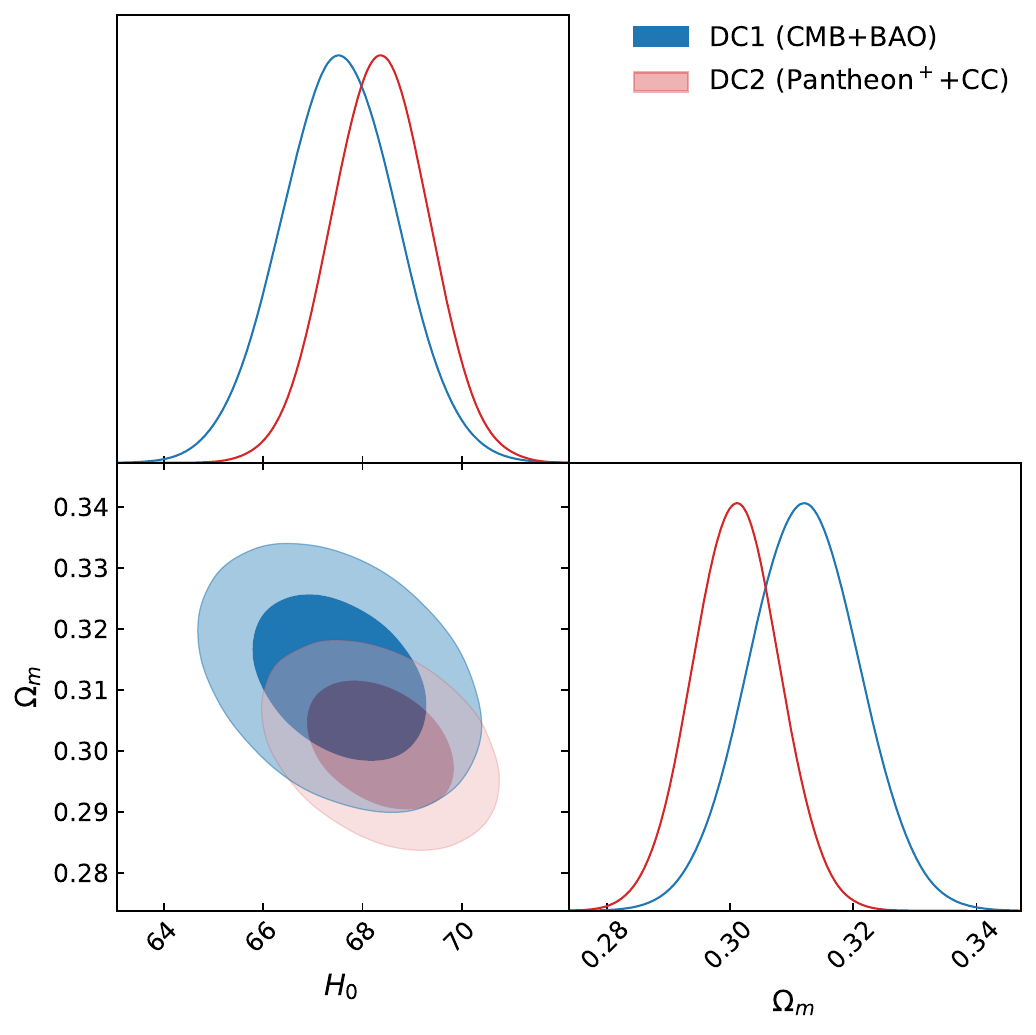}
	\caption{Marginalised one- and two-dimensional posterior distributions of $H_0$ and $\Omega_m$ for the $\Lambda$CDM model, from DC1 (CMB+BAO, orange) and DC2 (Pantheon$^+$+CC, blue). Contours show the $68\%$ and $95\%$ credible regions.}
	\label{fig:corner_lcdm}
\end{figure}

As the baseline against which the extended models are judged, $\Lambda$CDM gives $H_0 = 67.54 \pm 1.16$ and $\Omega_m = 0.306 \pm 0.005$ for DC1, and $H_0 = 68.36 \pm 0.98$ and $\Omega_m = 0.301 \pm 0.007$ for DC2 (Fig.~\ref{fig:corner_lcdm}). The two $H_0$ values differ by $0.82$~km\,s$^{-1}$\,Mpc$^{-1}$, about $0.5\sigma$ once the two uncertainties are combined in quadrature (Table~\ref{tab:h0_internal}); the $\Omega_m$ values agree at a similar level. The early- and late-Universe fits are therefore consistent with each other for this model, at the price of the largest $\chi^2_{\rm min}$ among the four models on both data combinations.

\subsection{$\Lambda$CDM with a free neutrino mass}
\label{subsec:lcdm_nu}

\begin{figure}[H]
	\centering
	\includegraphics[width=0.85\linewidth]{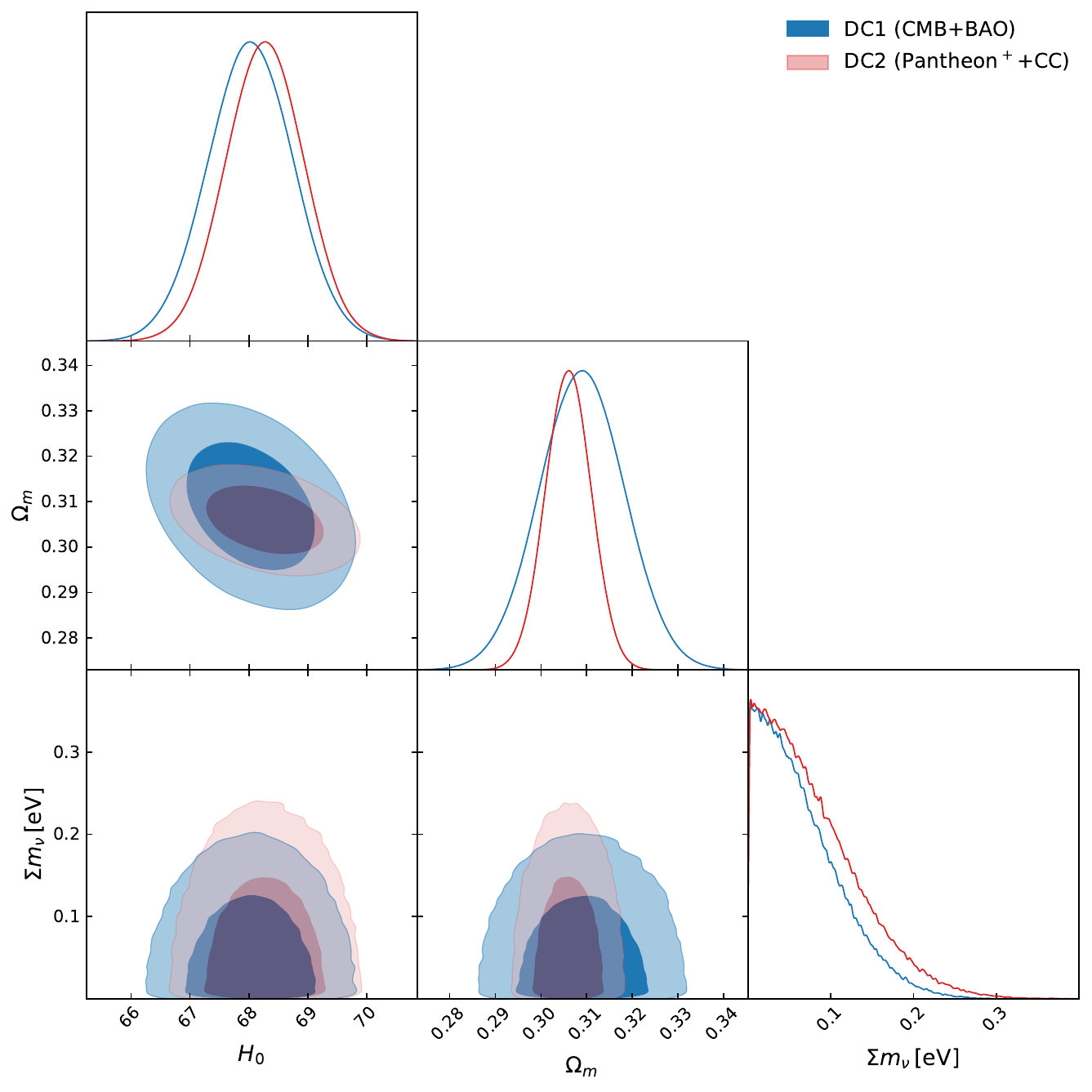}
	\caption{As in Fig.~\ref{fig:corner_lcdm}, for the $\Lambda$CDM model with massive neutrinos. The additional panel shows the marginalised posterior of $\sum m_\nu$, constrained from above on both data combinations.}
	\label{fig:corner_lcdm_withnu}
\end{figure}

Letting $\sum m_\nu$ float leaves $H_0$ close to its $\Lambda$CDM value on both combinations, $H_0 = 68.03 \pm 0.72$ (DC1) and $H_0 = 68.27 \pm 0.66$ (DC2), and brings the internal offset down to $0.3\sigma$ (Fig.~\ref{fig:corner_lcdm_withnu}, Table~\ref{tab:h0_internal}). The neutrino mass itself is only bounded from above, $\sum m_\nu < 0.16$~eV for DC1 and $< 0.19$~eV for DC2, consistent with a vanishing mass in both cases. The improvement in $\chi^2_{\rm min}$ relative to massless $\Lambda$CDM is modest on DC1 ($\Delta\chi^2 = 0.08$) and larger on DC2 ($\Delta\chi^2 = 16.02$), reflecting the additional constraining power of the larger sample.

\subsection{$f(Q,T)$ gravity without neutrino mass}
\label{subsec:fqt_nonu}

\begin{figure}[H]
	\centering
	\includegraphics[width=0.9\linewidth]{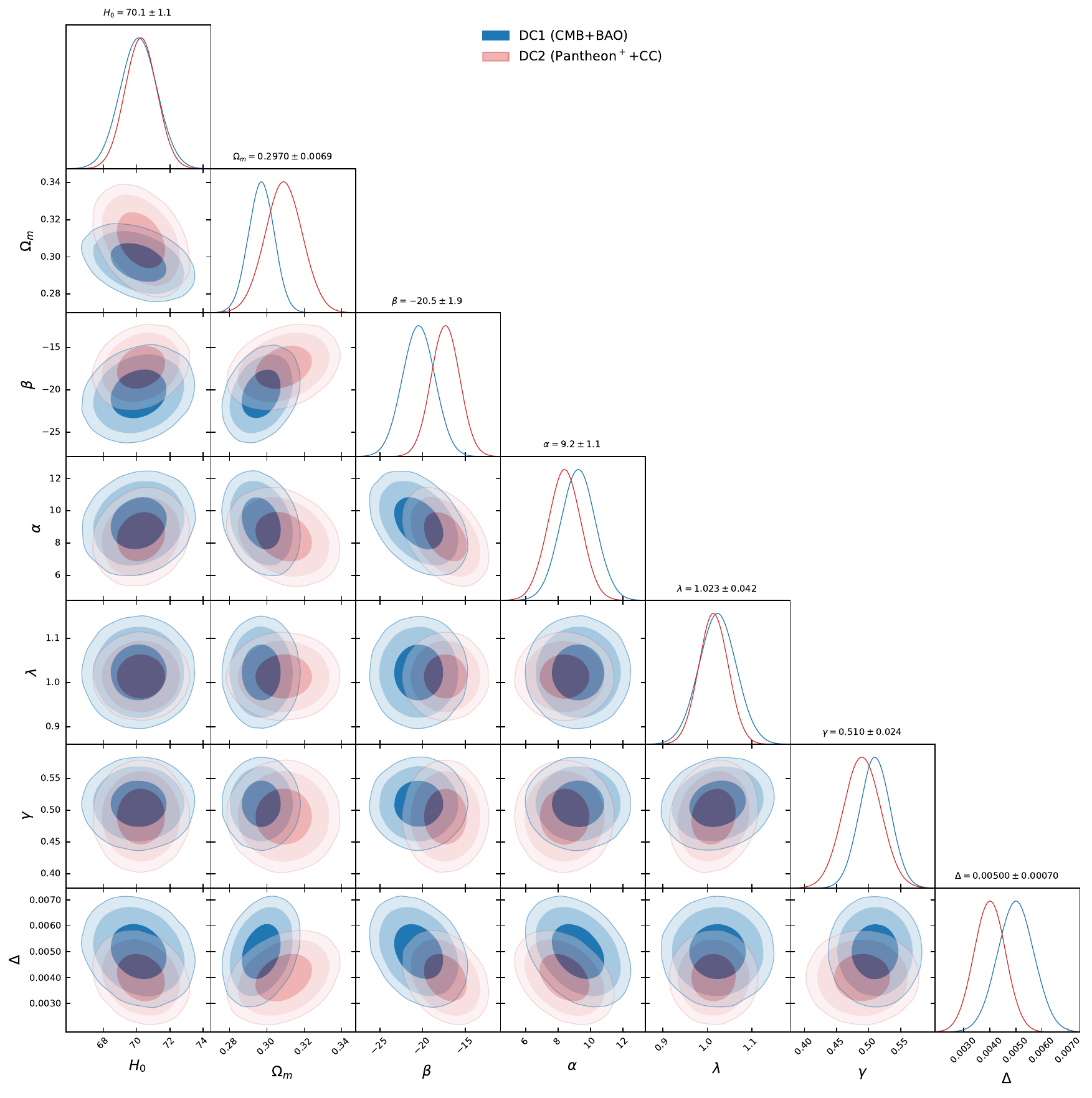}
	\caption{Marginalised posterior distributions of $H_0$, $\Omega_m$, and the $f(Q,T)$ parameters $\Delta$, $\gamma$, $\lambda$, $\alpha$, and $\beta$, without massive neutrinos, from DC1 (orange) and DC2 (blue). Contours show the $68\%$ and $95\%$ credible regions.}
	\label{fig:corner_fqt_nonu}
\end{figure}

Moving to $f(Q,T)$ gravity shifts $H_0$ further: $H_0 = 70.11 \pm 1.12$ (DC1) and $H_0 = 70.25 \pm 0.96$ (DC2), roughly $1.9$--$2.6$~km\,s$^{-1}$\,Mpc$^{-1}$ above the corresponding $\Lambda$CDM values (Fig.~\ref{fig:corner_fqt_nonu}). The internal DC1--DC2 offset in $H_0$ stays small, at $0.1\sigma$. Among the model-specific parameters, $\beta$ shows the largest early-to-late shift, $-20.47 \pm 1.89$ versus $-17.35 \pm 1.66$, an offset of $1.2\sigma$; $\alpha$, $\lambda$, and $\gamma$ agree to within $0.6\sigma$, and $\Delta$ to within $1.1\sigma$. None of the gravity-sector parameters differ between the two data combinations at a level that would suggest an inconsistency in the model.

\subsection{$f(Q,T)$ gravity with a free neutrino mass}
\label{subsec:fqt_nu}

\begin{figure}[H]
	\centering
	\includegraphics[width=0.9\linewidth]{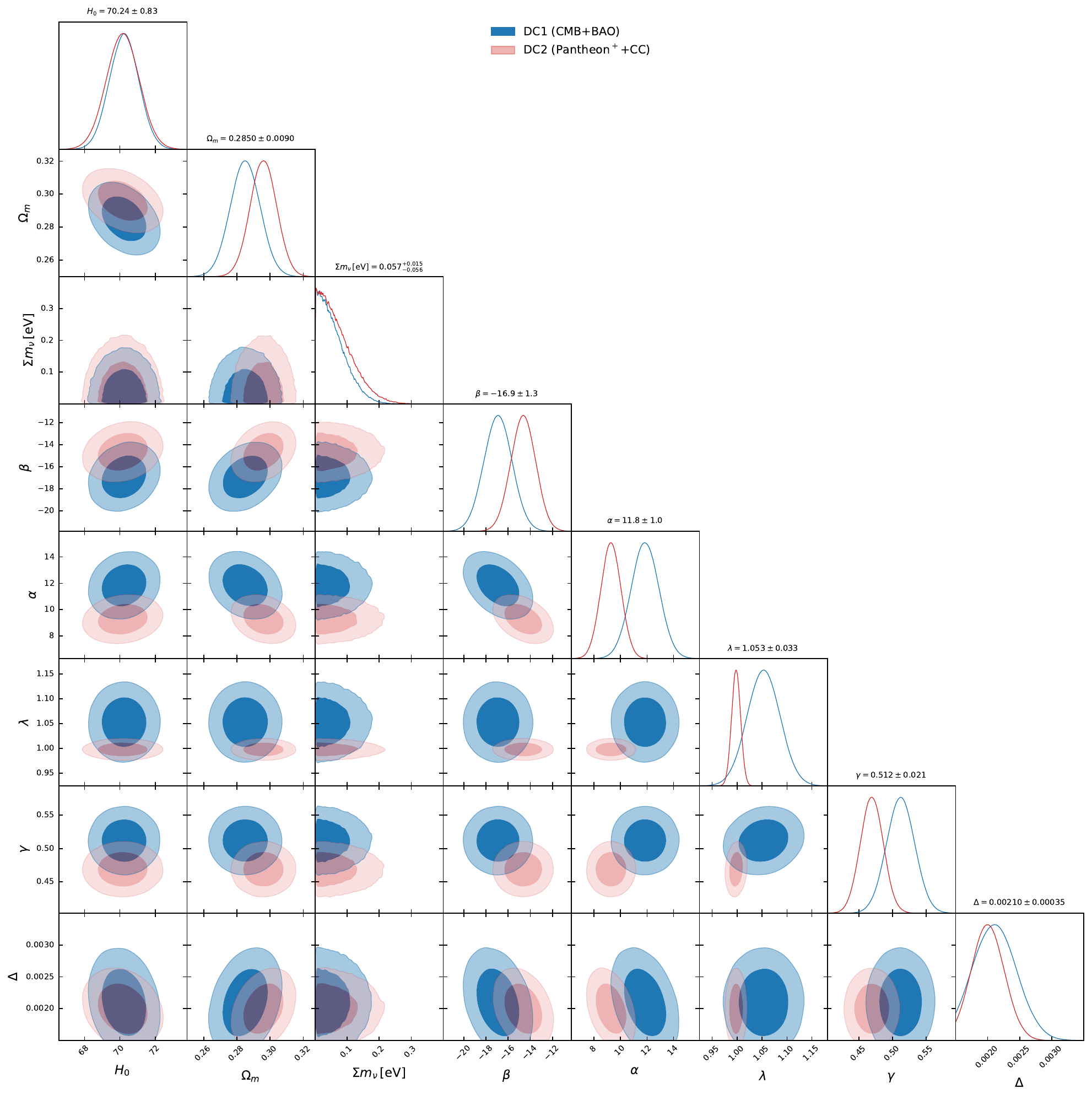}
	\caption{As in Fig.~\ref{fig:corner_fqt_nonu}, for the $f(Q,T)$ model with massive neutrinos included. The DC1 posterior for $\sum m_\nu$ peaks away from zero at $0.057^{+0.015}_{-0.056}$~eV, while both data combinations remain consistent with an upper limit.}
	\label{fig:corner_fqt_withnu}
\end{figure}

Adding $\sum m_\nu$ on top of $f(Q,T)$ gravity brings $H_0 = 70.24 \pm 0.83$ (DC1) and $H_0 = 70.16 \pm 0.93$ (DC2) to within $0.1\sigma$ of each other, the closest match among the four models (Fig.~\ref{fig:corner_fqt_withnu}, Table~\ref{tab:h0_internal}). The neutrino mass bound tightens slightly relative to $\Lambda$CDM, $\sum m_\nu < 0.14$~eV (DC1) and $< 0.17$~eV (DC2); the DC1 posterior peaks away from zero, at $\sum m_\nu = 0.057^{+0.015}_{-0.056}$~eV, though the bound remains an upper limit, not a detection. The gravity-sector parameters, by contrast, separate more between DC1 and DC2 than in the no-neutrino case: the offset rises to $1.3\sigma$ for $\beta$, $1.6\sigma$ for $\lambda$, $1.6\sigma$ for $\gamma$, and $2.0\sigma$ for $\alpha$, while it falls for $H_0$ and for $\Delta$ (to $0.2\sigma$). Adding the neutrino mass therefore improves the early-late agreement in the background parameters at the cost of a larger, though still not statistically significant, separation in the gravitational-sector parameters.

\subsection{Trend in $H_0$ across the four models}
\label{subsec:h0_trend}

Lined up together, the DC1 and DC2 values of $H_0$ increase in step, from $\Lambda$CDM ($67.5$--$68.4$) through $\Lambda$CDM with neutrinos ($68.0$--$68.3$) and $f(Q,T)$ without neutrinos ($70.1$--$70.3$) to $f(Q,T)$ with neutrinos ($70.2$--$70.2$), as summarised in Table~\ref{tab:h0_internal}. Within each model the early- and late-Universe values stay statistically consistent, at $0.1$--$0.5\sigma$; we note the trend without attaching a physical origin to it, since the present comparison addresses the internal consistency of each fit, not its interpretation.
\subsection{Model preference from AIC and BIC}
\label{subsec:aic_bic}

Consistency between DC1 and DC2 says nothing about which model the data prefer; for that we turn to the information criteria. For $k$ free parameters and $N$ data points, the two criteria are
\begin{equation}
	\mathrm{AIC} = \chi^2_{\rm min} + 2k, \qquad
	\mathrm{BIC} = \chi^2_{\rm min} + k\ln N,
	\label{eq:aic_bic}
\end{equation}
\cite{Akaike1974,Schwarz1978}, so that BIC penalises additional parameters more heavily than AIC once $\ln N > 2$, i.e.\ for $N > 8$ in DC1 and, more markedly, for $N=1733$ in DC2. $\Lambda$CDM is taken to have $k=3$ free parameters throughout this work, with one further parameter, $\Sigma m_\nu$, added for the neutrino extension. Table~\ref{tab:delta_ic} lists $\Delta\mathrm{AIC}$ and $\Delta\mathrm{BIC}$ relative to the $\Lambda$CDM baseline, which we interpret on the customary scale \cite{Jeffreys1961,KassRaftery1995,Liddle2007}: $|\Delta X| < 2$ inconclusive, $2 \le |\Delta X| < 6$ positive, $6 \le |\Delta X| < 10$ strong, $|\Delta X| \ge 10$ very strong evidence, with a negative sign favouring the extended model.

On DC1, the extended models are disfavoured relative to $\Lambda$CDM by both criteria, despite their lower $\chi^2_{\rm min}$: the neutrino extension of $\Lambda$CDM is inconclusive by AIC and positive by BIC, while the $f(Q,T)$ variants reach positive-to-strong preference for $\Lambda$CDM. This follows directly from Eq.~(\ref{eq:aic_bic}): with $N=16$, four or five extra parameters carry a penalty that the corresponding drop in $\chi^2_{\rm min}$ does not offset, so the small-sample result should be read as a statement about model complexity relative to the available data, not about the physical viability of the extended models.

On DC2, the same criteria favour the extended models by a wide margin, reaching very strong preference in every case. The two $f(Q,T)$ variants give the lowest AIC and BIC among the four models; the neutrino extension lowers AIC further ($\Delta\mathrm{AIC}=-53.02$ against $-47.67$) but leaves BIC close to the no-neutrino value ($-25.73$ against $-25.84$), since the extra parameter's $\ln N \approx 7.5$ penalty is close to the additional drop in $\chi^2_{\rm min}$ it buys. The DC1 and DC2 results together indicate that the preference for $f(Q,T)$ gravity found in the full analysis is carried by the large late-Universe sample, while the smaller early-Universe combination is, by construction, less able to support additional free parameters.

\begin{table}[H]
	\centering
	\caption{Cosmological parameter constraints for $\Lambda$CDM, $\Lambda$CDM (with neutrinos), $f(Q,T)$ (without neutrino), and $f(Q,T)$ (with neutrinos) models using DC1 ($N=16$) and DC2 ($N=1733$) datasets.}
	\label{tab:combined_results}
	\resizebox{\textwidth}{!}{
		\begin{tabular}{l cc| cc| cc| cc}
			\toprule
			& \multicolumn{2}{c}{\textbf{$\Lambda$CDM}} 
			& \multicolumn{2}{c}{\textbf{$\Lambda$CDM (with neutrinos)}} 
			& \multicolumn{2}{c}{\textbf{$f(Q,T)$ (without neutrino)}} 
			& \multicolumn{2}{c}{\textbf{$f(Q,T)$ (with neutrinos)}} \\
			\cmidrule(lr){2-3} \cmidrule(lr){4-5} \cmidrule(lr){6-7} \cmidrule(lr){8-9}
			Parameter & DC1 & DC2 & DC1 & DC2 & DC1 & DC2 & DC1 & DC2 \\
			\midrule
			$H_0$ & $67.54 \pm 1.16$ & $68.36 \pm 0.98$ & $68.03 \pm 0.72$ & $68.27 \pm 0.66$ & $70.11 \pm 1.12$ & $70.25 \pm 0.96$ & $70.24 \pm 0.83$ & $70.16 \pm 0.93$ \\
			$\Sigma m_\nu$ (eV) & - & - & $< 0.16$ & $< 0.19$ & - & - & $< 0.14$ & $< 0.17$ \\
			$\beta$ & - & - & - & - & $-20.47 \pm 1.89$ & $-17.35 \pm 1.66$ & $-16.92 \pm 1.28$ & $-14.64 \pm 1.12$ \\
			$\alpha$ & - & - & - & - & $9.23 \pm 1.07$ & $8.39 \pm 1.01$ & $11.85\pm 1.05$  & $9.28 \pm 0.76$ \\
			$\lambda$ & - & - & - & - & $1.023 \pm 0.042$ & $1.014 \pm 0.033$ & $1.053 \pm 0.033$ & $0.998 \pm 0.009$ \\
			$\gamma$ & - & - & - & - & $0.51 \pm 0.024$ & $0.49 \pm 0.029$ & $0.512 \pm 0.021$ & $0.469 \pm 0.017$ \\
			$\Omega_m$ & $0.306 \pm 0.005$ & $0.301 \pm 0.007$ & $0.309 \pm 0.0093$ & $0.306 \pm 0.005$ & $0.297 \pm 0.0069$ & $0.309 \pm 0.010$ & $0.285 \pm 0.009$ & $0.296 \pm 0.008$ \\
			$\Omega_\Lambda$ & $0.681 \pm 0.04$ & $0.685 \pm 0.07$ & $0.684 \pm 0.06$ & $0.677 \pm 0.012$ & - & - & - & - \\
			$\Delta$ & - & - & - & - & $0.004 \pm 0.0006$ & $0.004 \pm 0.0006$ & $0.0021 \pm 0.00035$ & $0.002 \pm 0.00026$ \\
			\midrule
			$\chi^2_{\rm min}$ & $12.06$ & $1589.33$ & $11.98$ & $1573.31$ & $9.07$ & $1533.66$ & $8.09$ & $1526.31$ \\
			$\rm AIC$ & $18.06$ & $1595.33$ & $19.98$ & $1581.31$ & $23.07$ & $1547.66$ & $24.09$ & $1542.31$ \\
			$\rm BIC$ & $20.38$ & $1611.70$ & $23.07$ & $1603.14$ & $28.48$ & $1585.86$ & $30.27$ & $1585.97$ \\
			\bottomrule
	\end{tabular}}
\end{table}

\begin{table}[H]
	\centering
	\caption{Internal consistency between the DC1 and DC2 determinations of $H_0$ for each model, expressed as $|\Delta H_0|/\sigma$ with $\sigma = \sqrt{\sigma_{\rm DC1}^2 + \sigma_{\rm DC2}^2}$.}
	\label{tab:h0_internal}
	\begin{tabular}{lccc}
		\toprule
		Model & $H_0$ (DC1) & $H_0$ (DC2) & Internal offset \\
		\midrule
		$\Lambda$CDM & $67.54 \pm 1.16$ & $68.36 \pm 0.98$ & $0.5\sigma$ \\
		$\Lambda$CDM (with $\nu$) & $68.03 \pm 0.72$ & $68.27 \pm 0.66$ & $0.3\sigma$ \\
		$f(Q,T)$ (without $\nu$) & $70.11 \pm 1.12$ & $70.25 \pm 0.96$ & $0.1\sigma$ \\
		$f(Q,T)$ (with $\nu$) & $70.24 \pm 0.83$ & $70.16 \pm 0.93$ & $0.1\sigma$ \\
		\bottomrule
	\end{tabular}
\end{table}

\begin{table}[H]
	\centering
	\caption{Information-criterion differences with respect to the baseline $\Lambda$CDM model, $\Delta X \equiv X_{\rm model} - X_{\Lambda\rm CDM}$, computed separately for the early-Universe (DC1) and late-Universe (DC2) combinations. A negative value indicates a fit favoured over $\Lambda$CDM by that criterion.}
	\label{tab:delta_ic}
	\begin{tabular}{l cc cc}
		\toprule
		& \multicolumn{2}{c}{DC1 (CMB+BAO)} & \multicolumn{2}{c}{DC2 (Pantheon$^+$+CC)} \\
		\cmidrule(lr){2-3}\cmidrule(lr){4-5}
		Model & $\Delta\mathrm{AIC}$ & $\Delta\mathrm{BIC}$ & $\Delta\mathrm{AIC}$ & $\Delta\mathrm{BIC}$ \\
		\midrule
		$\Lambda$CDM (with $\nu$) & $+1.92$ & $+2.69$ & $-14.02$ & $-8.56$ \\
		$f(Q,T)$ (without $\nu$) & $+5.01$ & $+8.10$ & $-47.67$ & $-25.84$ \\
		$f(Q,T)$ (with $\nu$) & $+6.03$ & $+9.89$ & $-53.02$ & $-25.73$ \\
		\bottomrule
	\end{tabular}
\end{table}

\subsection{Full dataset without neutrino mass: from CMB alone to CMB+all}
\label{sec:results-noNu}

Having compared the early- and late-Universe combinations separately, we now turn to the second step of the analysis: five overlapping combinations that move from the CMB distance priors on their own up to their combination with BAO, cosmic chronometers, and Pantheon$^+$. Table~\ref{tab:bestfit} lists the marginalized posterior constraints obtained when the neutrino sector is not varied, i.e.\ treated as effectively massless radiation. CMB+all denotes the combination of CMB+BAO+CC+Pantheon$^+$.

\begin{table}[H]
	\centering
	\caption{Marginalized posterior constraints on the cosmological and
		model parameters of the $f(Q,T)$--Barrow holographic dark energy
		model without massive neutrinos, for different combinations of CMB,
		cosmic chronometer (CC), and Pantheon+ supernova data. Values are
		posterior means with $1\sigma$ credible intervals; the last row
		gives the minimum $\chi^2$ for each combination.}
	\label{tab:bestfit}
	\begin{tabular}{lccccc}
		\toprule
		Parameter & CMB + CC & CMB + Pantheon$^+$ & CMB & CMB + BAO & CMB + all \\
		\midrule
		$H_0$        & $67.54 \pm 0.72$ & $70.80 \pm 0.68$ & $71.20 \pm 0.72$ & $70.11 \pm 1.12$ & $68.90 \pm 0.65$ \\
		$\Omega_b$   & $0.041 \pm 0.0017$ & $0.040 \pm 0.0020$ & $0.042 \pm 0.0010$ & $0.047 \pm 0.0009$ &$0.041 \pm 0.0007$ \\
		$\Omega_{cdm}$ & $0.261 \pm 0.009$ & $0.248 \pm 0.011$ & $0.246 \pm 0.006$ &$0.25 \pm 0.006$ & $0.262 \pm 0.005$ \\
		$\alpha$     & $15.1 \pm 2.1$  & $14.4 \pm 2.05$  & $10.56 \pm 1.10$ &$9.23 \pm 1.07$ & $9.5 \pm 0.85$ \\
		$\beta$      & $-16.9 \pm 2.1$ & $-14.8 \pm 1.7$  & $-15.66 \pm 1.35$ &$-20.47\pm1.89$ & $-18.57 \pm 1.15$ \\
		$\lambda$    & $1.060 \pm 0.075$ & $0.999 \pm 0.060$ & $1.084 \pm 0.040$ &$1.023 \pm 0.042$ & $0.998 \pm 0.030$ \\
		$\gamma$     & $0.508 \pm 0.040$ & $0.496 \pm 0.034$ & $0.546 \pm 0.022$ &$0.51 \pm 0.024$ & $0.499 \pm 0.018$ \\
		$\Delta$     & $0.0030 \pm 0.0007$ & $0.0029 \pm 0.0005$ & $0.0027 \pm 0.0004$ &$0.004 \pm 0.0006$ & $0.0020 \pm 0.0003$ \\
		\midrule
		$\chi^2$     & $17.754$ & $1519.091$ & $0.041$ &$9.07$ & $1543.41$ \\
		\bottomrule
	\end{tabular}
\end{table}

\textbf{Background parameters.} The Hubble constant recovered in this case
spans $H_0\simeq67$--$71$~km~s$^{-1}$~Mpc$^{-1}$ across the five
combinations, from $67.54\pm0.72$~km~s$^{-1}$~Mpc$^{-1}$ (CMB+CC) up to
$71.20\pm0.72$~km~s$^{-1}$~Mpc$^{-1}$ (CMB alone). This ordering seems
reasonable: CC data anchor $H(z)$ directly at intermediate redshift, whereas
CMB alone (or CMB with Pantheon+) is more sensitive to the shape than to the
absolute normalization of the expansion history. The baryon density is
comparatively stable across most combinations ($\Omega_b\simeq0.040$--$0.042$),
with a somewhat higher value for CMB+BAO ($\Omega_b=0.047\pm0.0009$); the
cold-dark-matter density, $\Omega_{cdm}\simeq0.246$--$0.262$, is likewise
stable across combinations and in line with standard expectations for the
matter content of the Universe.

\textbf{Coupling and holographic parameters.} The non-metricity--matter
coupling parameters $\alpha$ and $\beta$ depend on the data combination:
$\alpha$ ranges from $9.23\pm1.07$ (CMB+BAO) to $15.1\pm2.1$ (CMB+CC), while
$\beta$ stays negative throughout, from $-14.8\pm1.7$ (CMB+Pantheon+) to
$-20.47\pm1.89$ (CMB+BAO). We would not read too much into the precise
numerical values of $\alpha$ and $\beta$ on their own, given how they shift
between combinations; what looks more consistent across combinations is that
a non-trivial coupling ($\beta\neq0$) is preferred over the uncoupled limit
in every case. The Granda--Oliveros parameters are comparatively stable by
contrast: $\lambda$ stays close to unity ($0.998$--$1.084$) and $\gamma$
close to $0.5$ ($0.496$--$0.546$) throughout. The Barrow exponent $\Delta$ is
small and tightly constrained in every combination,
$\Delta\simeq0.002$--$0.004$, with no single data combination pulling away
from the rest; since $\Delta$ mainly governs the late-time evolution of
$\Omega_B$ [ Eq.~\eqref{eq:OmegaB-closure}], this agreement across
combinations indicates that the BHDE sector's late-time behaviour does not
depend strongly on which low-redshift probe anchors the fit.

\begin{figure}[H]
	\centering
	\includegraphics[width=\linewidth]{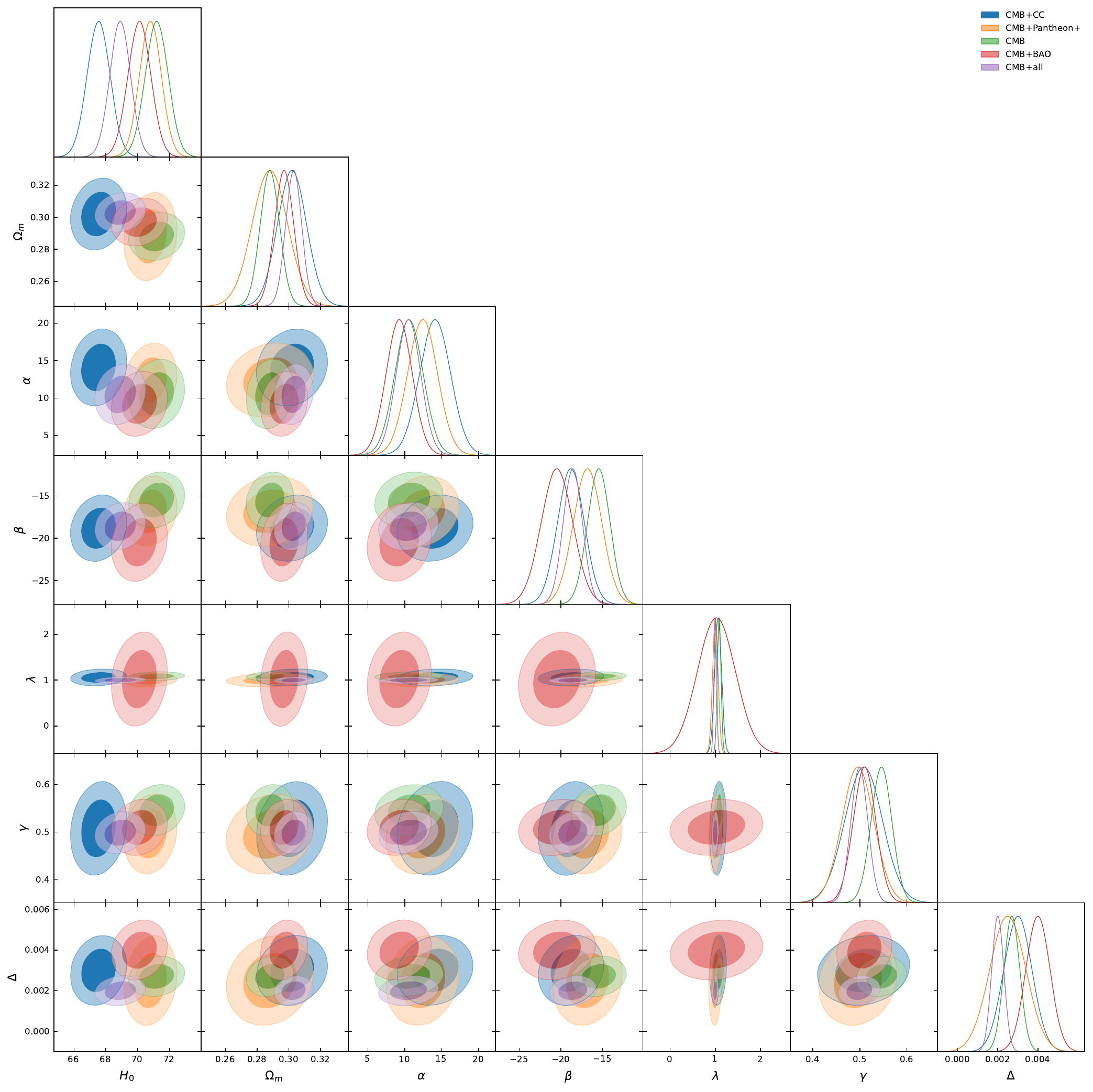}
	\caption{Marginalized one- and two-dimensional posterior distributions
		for the seven parameters of the $f(Q,T)$+BHDE model without massive
		neutrinos ($H_0$, $\alpha$, $\beta$, $\lambda$, $\gamma$, $\Delta$,
		$\Omega_m$), for the five data combinations of
		Table~\ref{tab:bestfit}.}
	\label{fig:triangle-noNu}
\end{figure}

Figure~\ref{fig:triangle-noNu} shows the corresponding marginalized posterior
distributions for all seven sampled parameters. The two-dimensional panels
are drawn in terms of the combined matter density
$\Omega_m\equiv\Omega_{cdm}+\Omega_b$ rather than the two components
separately; the individual values can be read off Table~\ref{tab:bestfit}.
We leave a detailed discussion of the parameter degeneracies visible in the
figure for future work, and limit ourselves here to noting that the five
data combinations occupy broadly overlapping regions of parameter space,
consistent with the posterior means already summarized above.

\subsection{Full dataset with a free neutrino mass}
\label{subsec:resultswithNu}

The natural next step is to let $\Sigma m_\nu$ vary in the same five
combinations, as we did earlier for the DC1/DC2 split. Table~\ref{tab:bestfit_nu}
repeats the analysis with $\Sigma m_\nu$ included as a free parameter, quoted
through its 95\% confidence-level upper bound.

\begin{table}[H]
	\centering
	\caption{Marginalized posterior constraints for the same model in
		the presence of massive neutrinos. The neutrino sector is quoted as
		the $95\%$ C.L.\ upper bound on $\sum m_\nu$; remaining parameters
		are posterior means with $1\sigma$ intervals. The last row gives
		the minimum $\chi^2$ for each combination.}
	\label{tab:bestfit_nu}
	\begin{tabular}{lccccc}
		\toprule
		Parameter & CMB + CC & CMB + Pantheon$^+$ & CMB & CMB + BAO & CMB + all \\
		\midrule
		$H_0$        & $68.21 \pm 1.01$ & $72.30 \pm 0.94$ & $70.60 \pm 0.98$ &$70.24 \pm 0.83$ &  $70.14 \pm 0.78$ \\
		$\Omega_b$   & $0.048 \pm 0.0018$ & $0.038 \pm 0.0022$ & $0.047 \pm 0.0011$ & $0.045 \pm 0.0010$ & $0.038 \pm 0.0009$ \\
		$\Omega_{cdm}$ & $0.259 \pm 0.010$ & $0.247 \pm 0.012$ & $0.246 \pm 0.007$ &  $0.241 \pm 0.008$ &$0.260 \pm 0.006$ \\
		$\sum m_\nu$ [eV] & $<0.15$ & $<0.16$ & $<0.22$ &  $<0.14$ &$<0.13$ \\
		$\alpha$     & $15.1 \pm 2.2$  & $14.4 \pm 2.06$  & $10.17 \pm 1.15$ & $11.85 \pm 1.05$ &$9.0 \pm 0.73$ \\
		$\beta$      & $-19.9 \pm 2.2$ & $-16.0 \pm 1.7$  & $-17.65 \pm 1.30$ &$-16.92 \pm 1.28$ & $-18.47 \pm 1.15$ \\
		$\lambda$    & $1.060 \pm 0.080$ & $0.998 \pm 0.062$ & $1.084 \pm 0.042$ & $1.053 \pm 0.033$ &$0.998 \pm 0.032$ \\
		$\gamma$     & $0.530 \pm 0.041$ & $0.499 \pm 0.035$ & $0.546 \pm 0.023$ &  $0.512 \pm 0.021$ & $0.499 \pm 0.019$ \\
		$\Delta$     & $0.0030 \pm 0.0006$ & $0.0050 \pm 0.0005$ & $0.0027 \pm 0.0004$ & $0.0021 \pm 0.00035$ &$0.0020 \pm 0.0003$ \\
		\midrule
		$\chi^2$     & $15.95$ & $1504.39$ & $0.03$ &$8.09$ & $1529.236$ \\
		\bottomrule
	\end{tabular}
\end{table}

\begin{figure}[H]
	\centering
	\includegraphics[width=\linewidth]{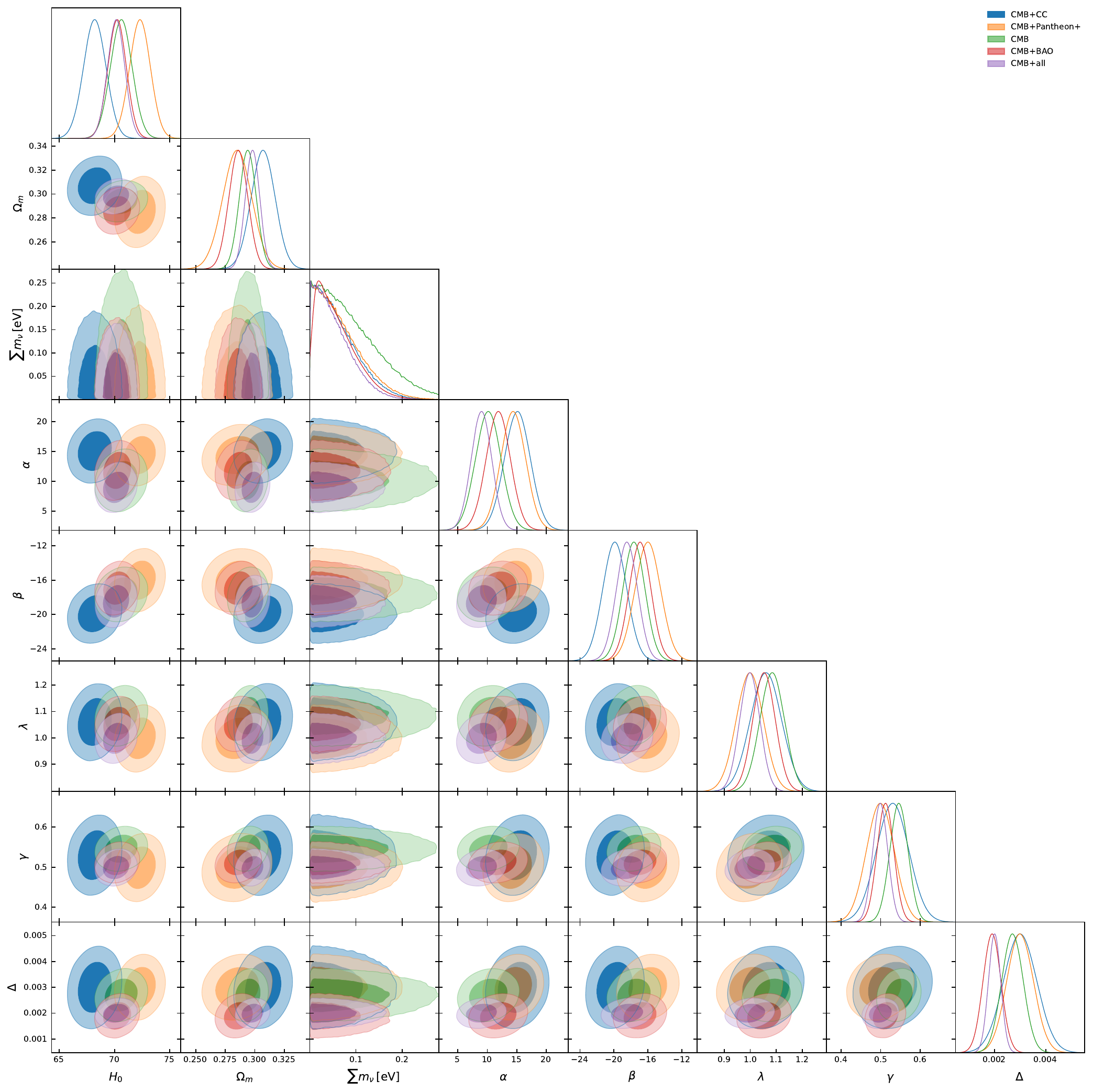}
	\caption{Marginalized one- and two-dimensional posterior distributions
		for the eight parameters of the $f(Q,T)$+BHDE model with massive
		neutrinos ($H_0$, $\alpha$, $\beta$, $\lambda$, $\gamma$, $\Delta$,
		$\Sigma m_\nu$, $\Omega_m$), for the same five data combinations as
		Table~\ref{tab:bestfit_nu}.}
	\label{fig:triangle-withNu}
\end{figure}

Figure~\ref{fig:triangle-withNu} shows the marginalized posteriors for all
eight sampled parameters, again with $\Omega_m\equiv\Omega_{cdm}+\Omega_b$
plotted in place of the two separate components.

\textbf{Shift relative to the massless case.} Allowing for a massive
neutrino sector shifts $H_0$ upward in most data combinations, though not
uniformly: from $67.54\pm0.72$ to $68.21\pm1.01$~km~s$^{-1}$~Mpc$^{-1}$ for
CMB+CC, and from $70.80\pm0.68$ to $72.30\pm0.94$~km~s$^{-1}$~Mpc$^{-1}$ for
CMB+Pantheon+. The CMB-only combination moves slightly the other way, from
$71.20\pm0.72$ to $70.60\pm0.98$~km~s$^{-1}$~Mpc$^{-1}$. The associated
uncertainties also grow once $\Sigma m_\nu$ is allowed to vary, so the
overall upward trend looks like a mild, data-driven tendency, not a strong
result on its own. The cold-dark-matter density shows at most a slight
decrease once neutrinos are added, and the size of the shift itself varies
with the data combination: it is essentially unchanged for CMB alone
($\Omega_{cdm}=0.246\to0.246$), a small decrease for CMB+CC
($0.261\to0.259$) and CMB+all ($0.262\to0.260$), and somewhat larger for
CMB+BAO ($0.25\to0.241$). This pattern is broadly consistent with part of
the non-relativistic matter budget being carried by the neutrino sector
instead of cold dark matter, though for most combinations here the shift is
small next to the quoted uncertainties, so we would not lean on it too
heavily as a standalone result. The coupling parameters $\alpha,\beta$ shift
by different amounts depending on the combination: $\alpha$ stays close to
its massless-neutrino value for CMB+CC and CMB+Pantheon+, while $\beta$
becomes somewhat more negative for CMB+CC ($-16.9\pm2.1\to-19.9\pm2.2$) and
CMB+Pantheon+ ($-14.8\pm1.7\to-16.0\pm1.7$). CMB+BAO shows the largest
change in either parameter, with $\alpha$ moving from $9.23\pm1.07$ to
$11.85\pm1.05$ and $\beta$ moving from $-20.47\pm1.89$ to $-16.92\pm1.28$;
CMB+all changes the least, in both $\alpha$ and $\beta$. The
Granda--Oliveros parameters $\lambda,\gamma$ stay close to their
massless-neutrino values throughout. This suggests that the non-metricity
and BHDE sectors are not strongly degenerate with the neutrino mass for most
combinations; CMB+BAO is the exception, where the two sectors respond to
each other more directly.

\textbf{Neutrino mass bound.} The 95\% C.L.\ upper limits obtained here,
$\Sigma m_\nu\lesssim0.13$--$0.22$~eV depending on the data combination, are
broadly comparable to, though somewhat looser than, typical bounds quoted in
the standard $\Lambda$CDM context. This may reflect the additional freedom
of the non-metricity and BHDE sectors partly absorbing the
neutrino-induced suppression of structure growth at the background level,
though confirming this would require extending the analysis beyond the
background equations used here.

\textbf{Internal consistency check.} As a basic sanity check, the minimum
$\chi^2$ decreases, rather than increases, in every combination once massive
neutrinos are introduced, e.g.\ from $1519.091$ to $1504.39$ for
CMB+Pantheon+, and from $17.754$ to $15.95$ for CMB+CC. Since the massless
case is formally the $\Sigma m_\nu\to0$ limit of the massive-neutrino model,
this is the behaviour expected of a properly converged fit, and it gives
some confidence that the two sets of chains are probing consistent regions
of parameter space.

\subsection{Comparison with $\Lambda$CDM across the full dataset}
\label{sec:aicbic}

As with the DC1/DC2 split, internal agreement between combinations does not
by itself tell us whether the data prefer the extended model; for that we
again turn to AIC and BIC, now for the five combinations of
Sec.~\ref{sec:results-noNu}. Table~\ref{tab:bestfit_lcdm} lists the
corresponding constraints for the standard $\Lambda$CDM model, fitted to the
same data combinations as Tables~\ref{tab:bestfit} and~\ref{tab:bestfit_nu}.

\begin{table}[H]
	\centering
	\caption{Marginalized posterior constraints on the cosmological
		parameters of the $\Lambda$CDM model for the same data
		combinations. Values are posterior means with $1\sigma$
		intervals; the last row gives the minimum $\chi^2$.}
	\label{tab:bestfit_lcdm}
	\begin{tabular}{lccccc}
		\toprule
		Parameter & CMB + CC & CMB + Pantheon$^+$ & CMB & CMB + BAO & CMB + all \\
		\midrule
		$H_0$            & $67.30 \pm 1.06$ & $72.84 \pm 0.95$ & $66.72 \pm 1.13$ & $67.54 \pm 1.16$ & $72.71 \pm 0.85$ \\
		$\Omega_m$       & $0.312 \pm 0.009$ & $0.297 \pm 0.011$ & $0.318 \pm 0.006$ & $0.306 \pm 0.005$ & $0.298 \pm 0.004$ \\
		$\Omega_\Lambda$ & $0.675 \pm 0.04$ & $0.673 \pm 0.05$ & $0.679 \pm 0.06$ & $0.681 \pm 0.04$ & $0.684 \pm 0.03$ \\
		\midrule
		$\chi^2$          & $29.55$ & $1554.24$ & $0.07$ & $12.06$ & $1591.78$ \\
		\bottomrule
	\end{tabular}
\end{table}

A direct comparison of the $\chi^2$ values across
Tables~\ref{tab:bestfit}--\ref{tab:bestfit_lcdm} is not, by itself, a
fair test, since the $f(Q,T)$+BHDE framework carries several more free
parameters than $\Lambda$CDM ($k=7$ without neutrinos, $k=8$ with,
versus $k=3$). We therefore also compute the Akaike and Bayesian
Information Criteria,
\begin{equation}
	\mathrm{AIC} = \chi^2_{\min} + 2k, \qquad
	\mathrm{BIC} = \chi^2_{\min} + k\ln N,
\end{equation}
where $N$ is the number of data points entering a given fit.
\paragraph{Information criteria.}
Table~\ref{tab:ic} collects $\chi^2_{\rm min}$, the number of free
parameters $k$, AIC ($=\chi^2_{\min}+2k$), and BIC
($=\chi^2_{\min}+k\ln N$) for the three models across the five data
combinations, with $\Delta\mathrm{AIC}$ and $\Delta\mathrm{BIC}$ given
relative to $\Lambda$CDM (negative values favor the extended model).
$N$ is fixed by the same effective sample sizes used throughout this
work (Sec.~\ref{sec:early_late_comparison}): $N_{\rm CMB}=3$
(compressed distance-prior points), $N_{\rm BAO}=13$, $N_{\rm CC}=32$,
and $N_{\rm Pantheon+}=1701$, summed as appropriate for each
combination.

\begin{table}[H]
	\centering
	\caption{Minimum $\chi^2$, number of free parameters $k$, AIC
		($=\chi^2_{\min}+2k$), and BIC ($=\chi^2_{\min}+k\ln N$) for the
		three models, using $N_{\rm CMB}=3$, $N_{\rm BAO}=13$,
		$N_{\rm CC}=32$, $N_{\rm Pantheon+}=1701$. $\Delta\mathrm{AIC}$ and
		$\Delta\mathrm{BIC}$ are relative to $\Lambda$CDM.}
	\label{tab:ic}
	\begin{tabular}{llccccccc}
		\toprule
		Combination ($N$) & Model & $k$ & $\chi^2_{\min}$ & AIC & $\Delta\mathrm{AIC}$ & BIC & $\Delta\mathrm{BIC}$ \\
		\midrule
		\multirow{3}{*}{CMB+CC ($N=35$)}
		& $\Lambda$CDM        & 3 & $29.55$   & $35.55$ & -- & $40.22$ & -- \\
		& $f(Q,T)$+BHDE       & 7 & $17.754$  & $31.75$ & $-3.80$ & $42.64$ & $+2.42$ \\
		& $f(Q,T)$+BHDE+$\nu$ & 8 & $15.95$   & $31.95$ & $-3.60$ & $44.39$ & $+4.17$ \\
		\midrule
		\multirow{3}{*}{CMB+Pantheon$^+$ ($N=1704$)}
		& $\Lambda$CDM        & 3 & $1554.24$  & $1560.24$ & -- & $1576.56$ & -- \\
		& $f(Q,T)$+BHDE       & 7 & $1519.091$ & $1533.09$ & $-27.15$ & $1571.18$ & $-5.38$ \\
		& $f(Q,T)$+BHDE+$\nu$ & 8 & $1504.39$  & $1520.39$ & $-39.85$ & $1563.92$ & $-12.64$ \\
		\midrule
		\multirow{3}{*}{CMB+all ($N=1749$)}
		& $\Lambda$CDM        & 3 & $1591.78$  & $1597.78$ & -- & $1614.16$ & -- \\
		& $f(Q,T)$+BHDE       & 7 & $1543.41$  & $1557.41$ & $-40.37$ & $1595.68$ & $-18.48$ \\
		& $f(Q,T)$+BHDE+$\nu$ & 8 & $1529.236$ & $1545.24$ & $-52.54$ & $1588.97$ & $-25.19$ \\
		\midrule
		\multirow{3}{*}{CMB ($N=3$)}
		& $\Lambda$CDM        & 3 & $0.07$  & $6.07$  & -- & $3.37$ & -- \\
		& $f(Q,T)$+BHDE       & 7 & $0.041$ & $14.04$ & $+7.97$ & $7.73$ & $+4.36$ \\
		& $f(Q,T)$+BHDE+$\nu$ & 8 & $0.03$  & $16.03$ & $+9.96$ & $8.82$ & $+5.45$ \\
		\midrule
		\multirow{3}{*}{CMB+BAO ($N=16$)}
		& $\Lambda$CDM        & 3 & $12.06$ & $18.06$ & -- & $20.38$ & -- \\
		& $f(Q,T)$+BHDE       & 7 & $9.07$  & $23.07$ & $+5.01$ & $28.48$ & $+8.10$ \\
		& $f(Q,T)$+BHDE+$\nu$ & 8 & $8.09$  & $24.09$ & $+6.03$ & $30.27$ & $+9.89$ \\
		\bottomrule
	\end{tabular}
\end{table}

For CMB+CC, AIC gives a small edge to the extended model
($\Delta\mathrm{AIC}=-3.80$ without massive neutrinos and $-3.60$ with
them), while BIC shows only marginal to positive evidence against it
on the Kass--Raftery scale ($\Delta\mathrm{BIC}=+2.42$ and $+4.17$).
For the combinations that include Pantheon+, both criteria favor the
extended model, with the preference strengthening once massive
neutrinos are included: under AIC this ranges from $-27.15$
(CMB+Pantheon+, no $\nu$) to $-52.54$ (CMB+all, with $\nu$), while
under BIC the preference is weaker but still positive to very strong,
from $-5.38$ to $-25.19$. The CMB-only combination shows the opposite
pattern under both criteria: with $\chi^2$ values this small, the
parameter penalty dominates and $\Lambda$CDM is favored
($\Delta\mathrm{AIC}\approx+8$ to $+10$, $\Delta\mathrm{BIC}=+4.36$ and
$+5.45$). CMB+BAO follows a similar, milder pattern under AIC
($\Delta\mathrm{AIC}\approx+5$ to $+6$) and reaches the ``strong''
category against the extended model under BIC
($\Delta\mathrm{BIC}=+8.10$ and $+9.89$). These two combinations
indicate that whatever advantage the extended model shows elsewhere
comes from the low-redshift supernova data specifically, not from a
generic feature of the fit.

Because BIC penalizes extra parameters more heavily than AIC once
$N>7$, the two criteria diverge for some combinations even though
both are computed from the same $\chi^2_{\rm min}$ values and the
same $N$. Both favor the extended model once Pantheon$^+$ is part of
the combination, most clearly for CMB+all, where $\Delta\mathrm{AIC}$
and $\Delta\mathrm{BIC}$ both reach their most negative values. For
CMB+CC, AIC favors the extended model while BIC favors $\Lambda$CDM,
both by a small margin; for CMB alone and for CMB+BAO, both criteria
favor $\Lambda$CDM, since the improvement in $\chi^2$ for these two
combinations does not offset the extra free parameters.

\section{The Hubble Tension in Light of the Present Model}
\label{sec:hubble-tension}

\begin{figure}[H]
	\centering
	\includegraphics[width=\linewidth]{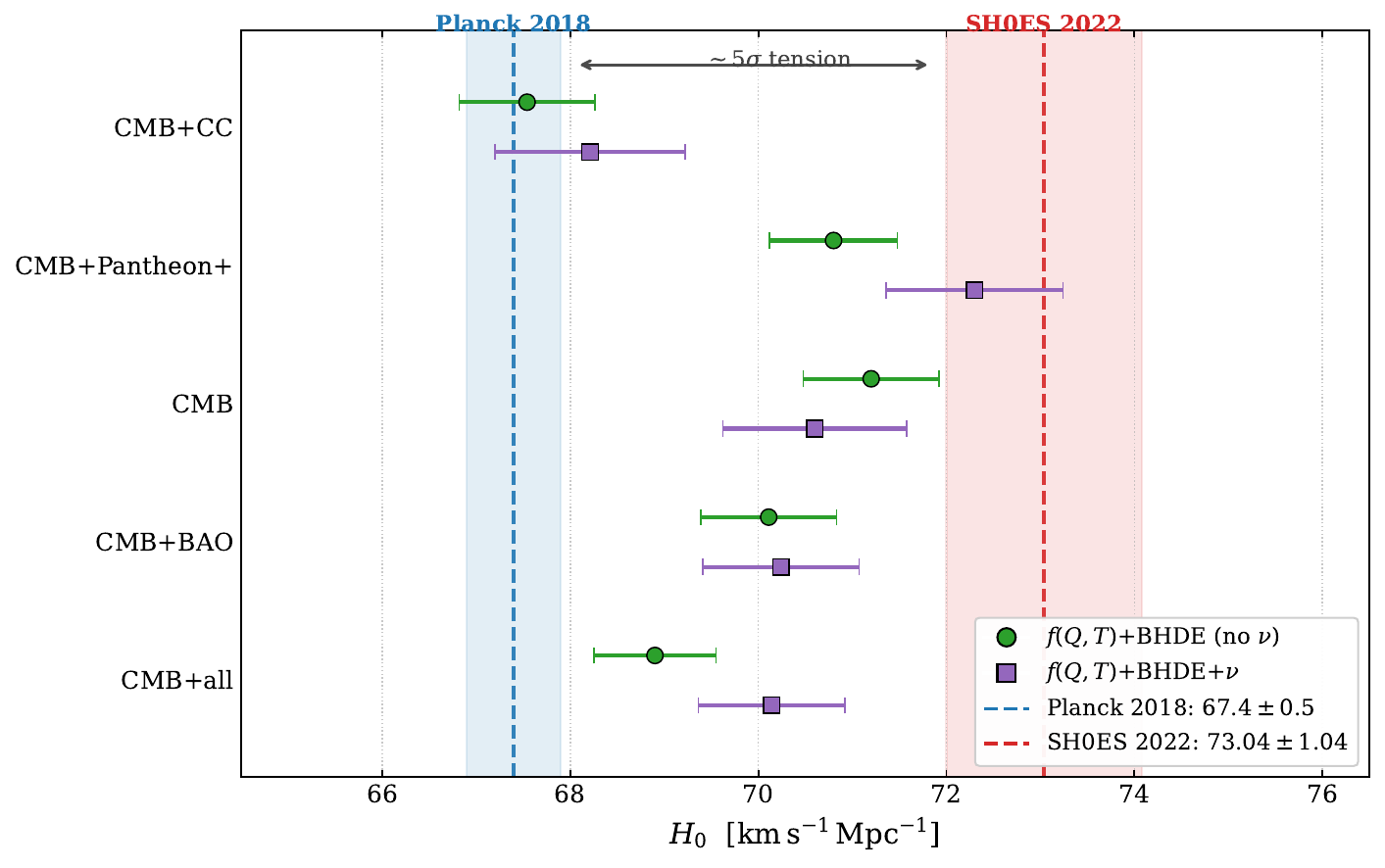}
	\caption{$H_0$ posterior means and $1\sigma$ errors for the
		$f(Q,T)$+BHDE model, with and without massive neutrinos, across the
		five data combinations of Tables~\ref{tab:bestfit} and
		\ref{tab:bestfit_nu}, compared with the Planck 2018
		($67.4\pm0.5$~km~s$^{-1}$~Mpc$^{-1}$) \cite{Planck2018} and SH0ES
		2022 ($73.04\pm1.04$~km~s$^{-1}$~Mpc$^{-1}$) \cite{Riess2022}
		determinations. The gap between the two shaded bands corresponds to
		the roughly $5\sigma$ tension quoted in the text; the exact tension
		for each point is given in Table~\ref{tab:H0tension}.}
	\label{fig:H0-tension}
\end{figure}

One of the most persistent challenges facing the standard $\Lambda$CDM model
is the so-called Hubble tension: the roughly 4--6$\sigma$ discrepancy between
the value of $H_0$ inferred from early-universe (CMB) observations and that
measured directly from the local distance ladder. The Planck 2018 analysis
of the CMB temperature and polarization power spectra, combined with
lensing, gives $H_0=67.36\pm0.54$~km~s$^{-1}$~Mpc$^{-1}$ within $\Lambda$CDM
\cite{Planck2018}, while the SH0ES collaboration's Cepheid-calibrated Type Ia
supernova measurement (commonly referred to as R22) reports
$H_0=73.04\pm1.04$~km~s$^{-1}$~Mpc$^{-1}$ \cite{Riess2022}. Our own
$\Lambda$CDM fits in Table~\ref{tab:bestfit_lcdm} reproduce this general
pattern: the CMB-only column gives $H_0=66.72\pm1.13$~km~s$^{-1}$~Mpc$^{-1}$,
close to the Planck value, while the Pantheon+-inclusive combinations give
$H_0\simeq72.7$--$72.8$~km~s$^{-1}$~Mpc$^{-1}$, close to R22. Figure~\ref{fig:H0-tension}
places the $H_0$ values from Tables~\ref{tab:bestfit}--\ref{tab:bestfit_lcdm}
alongside the Planck and R22 bands, and Table~\ref{tab:H0tension} lists, for
every combination and every model considered here, the corresponding
tension with both determinations, computed as
$|H_0-H_0^{\rm ref}|/\sqrt{\sigma^2+\sigma_{\rm ref}^2}$.

\begin{table}[H]
	\centering
	\caption{$H_0$ values from Tables~\ref{tab:bestfit}--\ref{tab:bestfit_lcdm}
		together with the corresponding tension, in units of $\sigma$, with
		the Planck 2018 ($67.36\pm0.54$~km~s$^{-1}$~Mpc$^{-1}$)
		\cite{Planck2018} and SH0ES R22 ($73.04\pm1.04$~km~s$^{-1}$~Mpc$^{-1}$)
		\cite{Riess2022} determinations of $H_0$. Tension is computed as
		$|H_0-H_0^{\rm ref}|/\sqrt{\sigma^2+\sigma_{\rm ref}^2}$.}
	\label{tab:H0tension}
	\begin{tabular}{llccc}
		\hline
		Combination & Model & $H_0$ [km s$^{-1}$ Mpc$^{-1}$] & Tension w/ Planck & Tension w/ R22 \\
		\hline
		\multirow{3}{*}{CMB+CC}
		& $\Lambda$CDM            & $67.30\pm1.06$ & $0.05\sigma$ & $3.87\sigma$ \\
		& $f(Q,T)$+BHDE           & $67.54\pm0.72$ & $0.20\sigma$ & $4.35\sigma$ \\
		& $f(Q,T)$+BHDE+$\nu$     & $68.21\pm1.01$ & $0.74\sigma$ & $3.33\sigma$ \\
		\hline
		\multirow{3}{*}{CMB+Pantheon+}
		& $\Lambda$CDM            & $72.84\pm0.95$ & $5.01\sigma$ & $0.14\sigma$ \\
		& $f(Q,T)$+BHDE           & $70.80\pm0.68$ & $3.96\sigma$ & $1.80\sigma$ \\
		& $f(Q,T)$+BHDE+$\nu$     & $72.30\pm0.94$ & $4.56\sigma$ & $0.53\sigma$ \\
		\hline
		\multirow{3}{*}{CMB+all}
		& $\Lambda$CDM            & $72.71\pm0.85$ & $5.31\sigma$ & $0.25\sigma$ \\
		& $f(Q,T)$+BHDE           & $68.90\pm0.65$ & $1.82\sigma$ & $3.38\sigma$ \\
		& $f(Q,T)$+BHDE+$\nu$     & $70.14\pm0.78$ & $2.93\sigma$ & $2.23\sigma$ \\
		\hline
		\multirow{3}{*}{CMB}
		& $\Lambda$CDM            & $66.72\pm1.13$ & $0.51\sigma$ & $4.12\sigma$ \\
		& $f(Q,T)$+BHDE           & $71.20\pm0.72$ & $4.27\sigma$ & $1.45\sigma$ \\
		& $f(Q,T)$+BHDE+$\nu$     & $70.60\pm0.98$ & $2.90\sigma$ & $1.71\sigma$ \\
		\hline
		\multirow{3}{*}{CMB+BAO}
		& $\Lambda$CDM            & $67.54\pm1.16$ & $0.14\sigma$ & $3.53\sigma$ \\
		& $f(Q,T)$+BHDE           & $70.11\pm1.12$ & $2.21\sigma$ & $1.92\sigma$ \\
		& $f(Q,T)$+BHDE+$\nu$     & $70.24\pm0.83$ & $2.91\sigma$ & $2.10\sigma$ \\
		\hline
	\end{tabular}
\end{table}

Within the $f(Q,T)$+BHDE framework we find a qualitatively similar spread,
shifted modestly toward higher values overall. Without massive neutrinos
(Table~\ref{tab:bestfit}), the CMB+CC combination gives
$H_0=67.54\pm0.72$~km~s$^{-1}$~Mpc$^{-1}$, in tension with Planck 2018 at
only $0.20\sigma$ and with R22 at $4.35\sigma$, while the CMB-only,
CMB+Pantheon+, and CMB+BAO combinations give $H_0=71.20\pm0.72$,
$70.80\pm0.68$, and $70.11\pm1.12$~km~s$^{-1}$~Mpc$^{-1}$, respectively --
intermediate between the Planck and R22 values, at $4.27\sigma$,
$3.96\sigma$, and $2.21\sigma$ from Planck and $1.45\sigma$, $1.80\sigma$,
and $1.92\sigma$ from R22. Including massive neutrinos
(Table~\ref{tab:bestfit_nu}) pushes several of these values higher still, up
to $H_0=72.30\pm0.94$~km~s$^{-1}$~Mpc$^{-1}$ for CMB+Pantheon+, bringing that
combination into closer -- though not exact -- agreement with R22
($0.53\sigma$, down from $1.80\sigma$ without neutrinos). The CMB-only and
CMB+BAO combinations move the other way, to $H_0=70.60\pm0.98$ and
$70.24\pm0.83$~km~s$^{-1}$~Mpc$^{-1}$, landing slightly closer to the Planck
value instead ($2.90\sigma$ and $2.91\sigma$ from Planck, versus
$4.27\sigma$ and $2.21\sigma$ without neutrinos).

These results are best read as a mild, data-combination-dependent easing of
the tension rather than a resolution of it: no single combination in
Tables~\ref{tab:bestfit}--\ref{tab:bestfit_nu} simultaneously matches both
the Planck and R22 central values, and the CMB+CC combination -- arguably
the one least affected by low-redshift systematics -- remains close to the
Planck value in both the massless- and massive-neutrino cases ($0.20\sigma$
and $0.74\sigma$, respectively). What the non-metricity coupling
($\alpha,\beta$) and the Barrow entropy correction ($\Delta$) appear to do
is introduce enough extra freedom in the late-time expansion history that
data combinations weighted toward low-redshift probes can accommodate
somewhat higher $H_0$ values than in $\Lambda$CDM, without requiring an
early-time modification of the sound horizon. This mechanism is broadly
similar in spirit -- though not in its physical origin -- to Early Dark
Energy or interacting dark-sector proposals discussed in the Introduction.
Neutrino-inclusive models of related kinds -- including neutrino-coupled $f(R)$ and Rastall gravity, Kaniadakis and Tsallis holographic dark energy, and CPL cosmology -- have likewise been reported to move $H_0$ toward the local value in their own analyses \cite{Yarahmadi2024Rastall,Yarahmadi2025Rastallnu,Yarahmadi2024Kaniadakis,Yarahmadi2025pertfR,Yarahmadi2025exploring,Yarahmadi2025noncanonical,Yarahmadi2026CPL,Yarahmadi2025BTPINN,Yarahmadi2025THDEPINN}. A quantitative comparison with those results lies outside the present background-level fits, because the models, datasets, and pipelines differ. It would also be worth exploring in future work whether a joint analysis
using the full CMB likelihood (rather than compressed distance priors, if
that is what has been used here) together with an explicit $H_0$ prior from
R22 could sharpen this conclusion.

\section{Conclusion}

We have examined a background-level cosmology built from $f(Q,T)=\alpha Q+\beta T$ gravity combined with a Barrow holographic dark energy density using the Granda-Oliveros cutoff, first with the neutrino sector fixed to zero mass and then with $\Sigma m_\nu$ treated as a free parameter. The model was fit to Pantheon+ supernovae, 32 cosmic chronometer measurements, DESI DR2 BAO, and Planck 2018 CMB distance priors using a modified \texttt{hi\_class} solver sampled with MontePython.

The fitted values of $H_0$ depend on which datasets enter the combination: cosmic-chronometer-dominated fits sit close to the Planck value, while combinations that include Pantheon+ move toward the SH0ES value, with no single combination matching both simultaneously. Allowing $\Sigma m_\nu$ to vary shifts $H_0$ upward in four of the five combinations, though the shift is modest relative to the resulting uncertainties, and only the CMB-only combination moves the other way. None of the combinations reduces the tension to an insignificant level: the closest approach to Planck remains CMB+CC ($0.20\sigma$ without neutrinos, $0.74\sigma$ with), and the closest approach to R22 is CMB+Pantheon+ with a massive neutrino ($0.53\sigma$). The 95\% bound on $\Sigma m_\nu$, between 0.13 and 0.22 eV across combinations, is close to though somewhat looser than typical $\Lambda$CDM bounds, consistent with part of the neutrino-induced suppression of structure growth being absorbed by the extra freedom in the $f(Q,T)$ coupling and the Barrow exponent $\Delta$; confirming this would require testing at the perturbation level rather than assuming it from the background fit alone.

The non-metricity coupling parameters $\alpha$ and $\beta$ shift by a few units between data combinations, with $\beta$ staying negative in every case; the Granda-Oliveros parameters $\lambda$ and $\gamma$ are more stable, remaining near 1 and 0.5 respectively, and the Barrow exponent $\Delta$ remains small throughout, at or just below 0.005 in every combination, indicating a very small departure from the standard entropy-area relation at the background level.

The AIC and BIC comparisons with $\Lambda$CDM do not agree on a single verdict for the extended model. Both criteria favor the extended model once Pantheon+ is part of the combination, and the preference strengthens once massive neutrinos are included; both criteria favor $\Lambda$CDM for CMB-only and CMB+BAO fits, where the improvement in $\chi^2$ is too small to offset the additional parameters. For CMB+CC, AIC gives a modest edge to the extended model while BIC gives a modest edge to $\Lambda$CDM. This pattern suggests that whatever preference the extended model shows here is tied to the low-redshift supernova sample specifically, rather than to a general feature of the fit, and any claim about its performance should specify which data combination it is based on.

Since the $H_0$ value used for Pantheon+ is calibrated through the Cepheid distance ladder, agreement with SH0ES in some combinations reflects agreement with that calibration rather than an independent resolution of the tension. The present analysis is also restricted to the background expansion: the sound horizon and decoupling redshift are held at their standard early-universe values, baryons and neutrinos are treated as pressureless matter at late times, and no perturbation-level quantity such as $f\sigma_8$ is computed.

\section{Acknowledgments}
I am incredibly grateful to the dear reviewer for the significant and valuable comments which caused this manuscript to improve considerably.

\end{document}